# Dispersive Elastodynamics of 1D Banded Materials and Structures: Design[*]

**Mahmoud I. Hussein**[†]
Post-Doctoral Researcher
Department of Mechanical Engineering
University of Michigan
Ann Arbor, MI 48103, USA

Currently:
Research Associate
Department of Engineering
University of Cambridge
Trumpington Street
Cambridge, CB2 1PZ, UK

Electronic Address: `mih21@cam.ac.uk`

**Gregory M. Hulbert**
Professor
Department of Mechanical Engineering
University of Michigan
2350 Hayward Street
2250 GG Brown Building
Ann Arbor, MI 48109, USA

**Richard A. Scott**
Professor
Department of Mechanical Engineering
The University of Michigan
1231 Beal Avenue
Auto Lab
Ann Arbor, MI 48109, USA

Number of Pages of Text:   33
Number of Figures:         27
Number of Tables:           0

---

# Abstract

Within periodic materials and structures, wave scattering and dispersion occur across constituent material interfaces leading to a banded frequency response. In an earlier paper, the elastodynamics of one-dimensional periodic materials and finite structures comprising these materials were examined with an emphasis on their frequency-dependent characteristics. In this work, a novel design paradigm is presented whereby periodic unit cells are designed for desired frequency band properties, and with appropriate scaling, these cells are used as building blocks for forming fully periodic or partially periodic structures with related dynamical characteristics. Through this multiscale dispersive design methodology, which is hierarchical and integrated, structures can be devised for effective vibration or shock isolation without needing to employ dissipative damping mechanisms. The speed of energy propagation in a designed structure can also be dictated through synthesis of the unit cells. Case studies are presented to demonstrate the effectiveness of the methodology for several applications. Results are given from sensitivity analyses that indicate a high level of robustness to geometric variation.

**Keywords:**  periodic materials, layered materials, phononic and photonic crystals, dispersion, band gaps, stop bands, pass bands, topology optimization, multiscale design

## 1. Introduction

Traditionally, overall passive vibration isolation using non-dissipative methods is achieved by employing one or more of the following techniques: (i) altering uniformly the material properties of a homogenous structure, (ii) altering the geometry of the structure, (iii) altering the way the structure is mounted or connected to adjacent structures, and/or (iv) reducing the excitation source. Occasionally these approaches are not satisfactory because a structure may be designed to satisfy other engineering criteria as well, and in such cases alterations could lead to violation of these other criteria. Another approach to vibration suppression involves using dissipative damping materials (e.g., viscoelastic materials) to absorb energy. However, energy absorbing materials typically do not have a high load-bearing capacity (among other disadvantages). Dissipative damping devices (e.g., shock absorbers) also are commonly used in practice, but these involve a damping mechanism that is exterior to the structure, which implies the need for proper joining as well as the need for occupying additional space. Other approaches include the concepts of sacrificial vibration absorbers and friction damping, but, as the names imply, an additional structure (or substructure) is required in each case. Furthermore, friction dampers often suffer from excessive wear and usually have limited life spans. The difficulties noted above also arise in the context of shock isolation.

This paper focuses on a type of structure that does not suffer the above disadvantages when employed for vibration/shock control, namely, *banded structures* (also known as band-gap structures). These are structures, composed of two or more constituent materials, that exhibit frequency-banded dynamical characteristics. Banded structures are fully or partially formed from periodically heterogeneous materials. Within these composite materials, wave scattering and dispersion occur across constituent material interfaces inducing mechanisms of constructive and destructive interference that lead to a banded frequency response. Within stop-band (or band-gap) frequency ranges, wave attenuation takes place and propagation is effectively prohibited. Conversely, within pass-band frequency ranges waves travel through the medium without attenuation. Periodic materials are alternatively referred to as phononic or sonic materials/crystals; the electromagnetic counterpart is known as photonic crystals.

Banded structures can be engineered to achieve vibration/shock isolation through passive control of the wave dispersion and scattering. Banded structures also can be designed to passively control speeds of wave propagation and hence energy flux velocities [1]. Thus, a major advantage is that wave attenuation is forced to take place within the structure itself requiring no additional components. This implies that the structure can be *designed totally from within*, that is, the design variables could comprise only the properties of constituent materials and their topological distribution within the structure. Also, no special alterations to the geometry or to the connectivity of the structure to other parts are required, since the dynamical characteristics are controlled primarily through the choice and distribution of the constituent materials. Since the key factor in choosing the materials is the ratio of their elastic moduli and densities, banded structures can be made from stiff and high strength materials[1] so as not to sacrifice load-bearing capacity, and can be manufactured using heat and chemical resistant materials to operate in a wide range of environments.

Utilization of stop-band and pass-band characteristics for material design has been considered in the literature. For example, Sigmund and Jensen [2] used topology optimization techniques to synthesize a two-dimensional periodic unit cell (representing an infinite periodic material) exhibiting a band gap (between two selected branches) with maximized width and minimized central frequency. Earlier in the related field of electromagnetics (i.e., photonic crystals), Cox and Dobson [3-4] also optimized unit cells for a band gap with maximum width and centered at a given frequency. However, none of these authors considered designing a unit cell with the objective of maximizing the wave attenuation strength at a particular frequency or maximizing the presence of band gaps over a wide frequency range that exceeds the maximum attainable width of a single band gap.

Vibration control of finite (i.e., bounded) periodic structures has also been investigated in recent years. For example, Ruzzene and Baz [5] studied periodic composites controlled with shape memory inserts, Baz [6] studied the active vibration control of lumped mass periodic structures, Ruzzene et al. [7] studied periodic cellular grid structures and how frequency-dependent phase constant surfaces could be used to control the directions of wave propagation, and Richards and Pines [8] applied stop-band/pass-band concepts associated with geometric periodicity to the reduction of gear mesh vibration. Recently, Diaz et al. [9] studied and designed band-gap grid structures in which non-structural masses have been included as design variables. Burger et al. [10] provided a review of advances on inverse problem techniques for the design of photonic crystals (which is relevant to phononic crystals as well). On another track, direct topology optimization of the bounded structure has been extensively studied for the objective of designing composite structures with improved vibration characteristics. For example, Díaz and Kikuchi [11] addressed eigenvalue problems, and Sigmund and Jensen [2] designed waveguides. Other studies that focused on topology or shape optimization in structural dynamics problems include those of Ma et al. [12], Min et al. [13], Zhao et al. [14], and Lai and Ananthasuresh [15].

---

[1] This, however, comes at the expense of increasing the minimum allowable operational frequency ranges or minimum structure dimensions for feasible designs (see Section 3.3).

Phenomena associated with banded dynamics were analyzed earlier in the context of both infinite periodic materials and finite periodic structures [16,17]. It was shown that with a few cells, the dynamic behavior of a periodic structure qualitatively matches that of the constituent periodic material. It was shown that with knowledge of the attenuation constant corresponding to a stop-band excitation frequency, the strength of attenuation in the structure could approximately be predicted as a function of the number of cells employed. The present paper builds upon this foundation by presenting a novel methodology whereby periodic materials are designed with desired frequency band properties, and, with appropriate scaling, are used as building blocks to form dispersive structures with related frequency-dependent dynamical characteristics.

This strategy provides promising opportunities for the design of engineering structures for a wide range of applications. For example, banded structures could be designed to act as non-dissipative (or dissipative if additional attenuation is required) vibration isolation devices, or joints, which are capable of reducing vibration transmission at a particular frequency, or at several specified frequencies, or across a single wide range of frequencies (preliminary results for material design are given in Ref. [18] and for material-structure design in Ref. [16,19,20]). Furthermore, banded structures could be employed as components in their own right serving separate engineering functions other than vibration isolation, yet exhibiting specific frequency-dependent dispersive dynamical characteristics.

The proposed design methodology for banded structures is simple and physically driven since it stems from an understanding of the dynamics of periodic materials and structures. Furthermore, the implementation of the methodology is efficient because it effectively reduces the design space of the entire structure to that of a single or a few different unit cells (at the material level) coupled with a coarser design space that is to be filled with the scaled cell building blocks (at the structural level). This implies that it is sufficient to conduct optimization (or an exhaustive search for the case of simple models) on only a single or a few unit cells as opposed to the structure as a whole, followed by a rather coarse design space search at the structural level.

There are four objectives addressed in this paper. The first is the introduction of the *multiscale dispersive design methodology*. The material design focuses solely on maximizing wave attenuation. The structural design involves a simple approach in which the entire bounded domain is composed of synthesized cells of a single design, or more advanced approaches in which generally more than one unit cell design are used for cells that fully, or possibly only partially, occupy the bounded domain. The second objective is the demonstration of the applicability of the proposed methodology to cases of both vibration and shock isolation (the former is due to harmonic excitation and the latter is due to transient broad-band pulse excitation). For the problem of vibration reduction, the case of multiple excitations at different locations and frequencies is also addressed. The third objective is the demonstration of how the speed of energy propagation in a designed banded structure could be dictated through synthesis of the unit cell. Conducting sensitivity analyses aimed at investigating the degree of robustness of periodic materials to disorder in the thicknesses of the layers is the fourth objective. All models considered in this work are one dimensional (1D). Generalization of the methodology to 2D materials and structures is presented in Refs. [16,19,20].

In the next section, the governing equations for linear elastodynamics of 1D banded materials and structures are given and solution methods outlined. The proposed multiscale dispersive design methodology objectives, description, implementation approaches are presented in Section 3. In Section 4, the application of the methodology to the cases of vibration isolation (due to single and multiple excitations), shock isolation, and control of energy propagation speed is covered. An analysis of the sensitivity of optimal material designs to geometric disorder is presented in Section 5. Finally, conclusions are drawn in Section 6.

## 2. Linear Elastodynamics of 1D Banded Materials and Structures

For the ease of exposition, longitudinal wave propagation in 1D heterogeneous solids is considered. The governing equation of motion is given by

$$\rho \ddot{u} = \sigma,_x + f, \qquad (1)$$

where $\rho = \rho(x)$, $f = f(x,t)$, $\sigma = \sigma(x,t)$, and $u = u(x,t)$ denote density, external force, stress, and displacement, respectively. The position coordinate and time are respectively denoted by $x$ and $t$. The $x$ subscript denotes partial differentiation with respect to position, while a superposed dot denotes partial differentiation with respect to time. In this paper, only linear elastic materials are assumed, i.e.,

$$\sigma = E u,_x, \qquad (2)$$

where $E = E(x)$ is the Young's modulus. The following subsections define the mathematical models and solution methods for the various problems considered in this work.

### 2.1 Periodic Materials

Due to symmetry, a model of a periodic medium can be reduced to that of a single unit cell with periodic boundary conditions. The length of a cell is denoted $d$ and consists of $n$ layers of alternating material property (in this work only two material phases are considered for simplicity). The $j$th layer has thickness $d^{(j)}$ and propagation speed $c^{(j)} = \sqrt{E^{(j)}/\rho^{(j)}}$ (the superscript refers to property of layer $j$). The interfaces between the layers are assumed to be ideal. Thus, the boundary conditions that must be satisfied at the layer interfaces are (i) continuity of displacement, $u$, and (ii) continuity of stress, $\sigma$. No external forces are permitted, i.e., $f = 0$ in Eq. (1).

The displacement solution $u = u(x,t)$ in the $j$th layer is written as a superposition of forward and backward traveling waves with harmonic time dependence:

$$u(x,t) = [A_+^{(j)} e^{ik^{(j)}x} + A_-^{(j)} e^{-ik^{(j)}x}] \times e^{-i\omega t}, \tag{3}$$

where $i = \sqrt{-1}$, $k^{(j)} = \omega/c^{(j)}$ and $\omega$ is the temporal frequency. Due to spatial periodicity, a Floquet solution is sought, i.e.,

$$\begin{bmatrix} u(x+d) \\ \sigma(x+d) \end{bmatrix} = e^{ikd} \begin{bmatrix} u(x) \\ \sigma(x) \end{bmatrix}, \tag{4}$$

where $k$ is the Floquet wavenumber (Ref. [21]). The modeling of the unit cell layers is formulated using the Transfer Matrix method (as described in Refs. [16,17,22,23]). The frequency spectrum is subsequently obtained by solving the dispersion relation, which is expressed as the following eigenvalue problem:

$$\mathbf{T}(\omega) \begin{bmatrix} u(x) \\ \sigma(x) \end{bmatrix}_{x=x^{1L}} = \lambda \begin{bmatrix} u(x) \\ \sigma(x) \end{bmatrix}_{x=x^{1L}}, \tag{5}$$

where $\lambda = \exp(ikd)$ is a complex eigenvalue, $[u(x), \sigma(x)]_{x=x^{1L}}^{\mathrm{T}}$ is a complex eigenvector, and $x = x^{1L}$ is the position of the left end of the unit cell. The frequency-dependent cumulative transfer matrix for the unit cell is defined as

$$\mathbf{T} = \mathbf{T}_n \mathbf{T}_{n-1} \cdots \mathbf{T}_1, \tag{6}$$

where, defining $Z^{(j)} = \rho^{(j)} c^{(j)^2} k^{(j)}$, the transfer matrix for layer $j$ is

$$\mathbf{T}_j = \begin{bmatrix} \cos(k^{(j)} d^{(j)}) & (1/Z^{(j)}) \sin(k^{(j)} d^{(j)}) \\ -Z^{(j)} \sin(k^{(j)} d^{(j)}) & \cos(k^{(j)} d^{(j)}) \end{bmatrix}. \tag{7}$$

### 2.2 Periodic or Partially Periodic Structures

For wave propagation analyses, periodic or partially periodic structures in this study are modeled using Eqs. (1)-(2) along with a problem-specific set of initial and boundary conditions.

A standard finite element method and time integration scheme are used for obtaining solutions, the details of which are specified for each example problem.

For some cases in this work, the steady-state forced vibration response of a finite structure is also sought. Here, the governing equation again is given by Eqs. (1)-(2), but now it is assumed that the applied force is harmonic with frequency $\omega^*$ and has a complex amplitude of $F(x, \omega^*)$, that is,

$$f(x,t) = F(x,\omega^*)e^{i\omega^* t}. \tag{8}$$

Hence the steady-state forced vibration response is

$$u(x,t) = U(x,\omega^*)e^{i\omega^* t}, \tag{9}$$

where $U$ is also a complex coefficient carrying frequency-dependent magnitude and phase information. Substituting Eqs. (8)-(9) into Eqs. (1)-(2) gives

$$-\omega^{*2}\rho U = EU_{,xx} + F. \tag{10}$$

With an appropriate set of boundary conditions, Eq. (10) forms a boundary value problem which is also solved using finite elements.

## 3. Multiscale Dispersive Design Methodology
### 3.1 Objectives

The proposed methodology is capable of satisfying one or more of the following objectives:

(i) attainment of low level vibration everywhere in the structure when it is subjected to harmonic excitation at any given location(s) (vibration suppression),

(ii) prevention of waves associated with harmonic vibration loads from propagating from one part of the structure to another (vibration isolation),

(iii) prevention of waves associated with shock loads from propagating from one part of the structure to another (shock isolation),

(iv) control of the speed of energy propagation in the structure.

The engineering relevance of the first three objectives is obvious. The fourth objective provides a capability to control the timing of energy propagation in a structure, which can be utilized in new structural vibration control strategies.

## 3.2 Description

The *multiscale dispersive design (MDD)* methodology utilizes the facts that (i) varying the configuration of the unit cell of a periodic material allows one to control the frequency band structure, i.e., the size and location of stop and pass bands [2-4,16,18,20,24], and (ii) with a few cells stacked adjacent to each other, the dispersive dynamic behavior in a structure qualitatively matches that of the constituent periodic material [16-17]. The methodology stipulates that the topology of a periodic composite material, or more than one composite material, is synthesized, and with appropriate scaling these designed materials are used to form a bounded structure with frequency-banded dynamical characteristics that correlate with those of the materials. The spatial allocation of the periodic materials, or groups of stacked same-type cells, in the structure is guided by the location of sources of excitation and the targeted dynamical response, both locally and of the structure as a whole. Typically, a larger length scale is associated with the structure. The underlying material-structure design process is inherently hierarchical, and in some cases it is integrated.

Within the framework of the proposed methodology, a variety of design techniques are possible to enhance the performance characteristics of the designed structures and/or to reduce the volume of periodic materials employed. Fig. 1 shows a schematic of examples of synthesized banded materials (building blocks) and examples of banded structures that are designed using the banded materials.

## 3.3 Implementation

The specific formulation and solution technique for both the unit cell and structure design problems are independent from the core MDD methodology. Each of the design problems could be solved by a variety of methods ranging from a trial-and-error approach to automated optimization. In Section 3.3.1, a candidate formulation for the material design problem is presented, and in Section 3.3.2, two sets of guidelines for the material-structure design problem are presented.

### 3.3.1 A Formulation for Material (Unit Cell) Design

Attention is restricted to a two material phase unit cell. Assume that a unit cell is composed of $n$ layers of either relatively stiff (fiber) material or relatively compliant (matrix) material (distinguished with the subscripts 'f' and 'm', respectively). The layers alternate in type and their thicknesses are expressed in vector form, $\bm{d} = (d^{(1)}, d^{(2)}, \ldots, d^{(n)})$. For practical considerations, assume that the layer thickness cannot be smaller than $a$. Also, the total length of the cell is set to unity, resulting from non-dimensionalization of the model parameters. On this basis, the material layout can be formulated in terms of a vector of binary variables $\bm{b} = (b_1, b_2, \ldots, b_l)$ (with a constant dimension $l$) by dividing the unit cell into $l$ divisions (or "slots") each of which can be filled with either of the two materials [18,24]:

$$b_i = \begin{cases} 0 & \text{if slot } i \text{ is filled with material 'f'} \\ 1 & \text{if slot } i \text{ is filled with matrix 'm'} \end{cases}. \tag{11}$$

Adjacent slots filled with the same material are contiguous and hence form a single homogenous cell layer. This description naturally enables a binary encoding scheme for representation of the configuration of the unit cell and hence the periodic material.

The total number of layers is

$$n = \sum_{i=1}^{l-1} s_i + 1, \quad (12)$$

where $s = (s_1, \ldots, s_{l-1})$, $s_i = \text{XOR}(b_i, b_{i+1})$. The thickness $d^{(j)}$ of layer $j$ can take only discrete values with the multiple of $1/l$, which can be expressed as:

$$d^{(j)} = (q_j - q_{j-1}) \times \frac{1}{l}, \quad (13)$$

where $q = (q_0, q_1, \ldots, q_n)$ are the indices of $s$ with $s_i = 1$, sorted in the ascending order with $q_0 = 0$ and $q_n = l$. For example, let $l = 10$; the 1st and 3rd layers are made of the compliant material and are, respectively, 3 and 1 division(s) thick; the 2nd and 4th layers are composed of the stiff material and are each 3 divisions thick. For this configuration, $b = (1,1,1,0,0,0,1,0,0,0)$. Thus, $s = (0,0,1,0,0,1,1,0,0)$ and $q = (0,3,6,7,10)$, and hence $n = 4$ and $d = (0.3,0.3,0.1,0.3)$. Permitted by periodicity, the first layer can always be taken as a particular type (stiff or compliant) and the remaining layers alternate in type. In terms of $l$, the total number of layers is

$$n = \dim(q)-1. \quad (14)$$

With this form of representation, the above example unit cell is encoded as '1110001000'. This unit cell is shown in the top left of Fig. 1. Another example unit cell, represented as '1111100000', is shown in the bottom left of Fig. 1. In addition to facilitating computer programming for topology optimization of the unit cell, this encoding scheme is convenient for providing a unique labeling system for the unit cells, which could conveniently be referred to as *binary code unit cells*.

Using $b$ as an independent design variable, the optimization problem can now be cast as a zero-one integer programming formulation [24]:

$$\text{Minimize:} \quad g = \text{Pen}(n, \boldsymbol{d}), \tag{15}$$

$$\text{Subject to:} \quad n = \sum_{i=1}^{l-1} s_i + 1, \; s_i = \text{XOR}(b_i, b_{i+1}) \text{ for } i = 1, 2, \ldots, l\text{-}1, \tag{16}$$

$$d^{(j)} = (q_j - q_{j-1}) \times \frac{1}{l}, \; q_j \text{ for } j = 0, \ldots, n, \tag{17}$$

$$d^{(j)} \geq a \text{ for } j = 1, 2, \ldots, n, \tag{18}$$

$$2 \leq n \leq n_{\max}, \; n \text{ is an even number}, \tag{19}$$

$$\boldsymbol{b} \in \{0,1\}^l. \tag{20}$$

The form of $g$ depends on the design application; examples are given in Section 4. Increasing the number of "bits" (value of $l$) in the unit cell leads to a better optimal design (up to a limit as discussed in Ref. [24]). When $l$ is large, advanced gradient-based or heuristic techniques (such as genetic algorithms) should be utilized for solving Eqs. (15)-(20). In this work, however, the value of $l$ is kept relatively small enabling an exhaustive search to be conducted across all possible binary combinations (i.e., all possible cell configurations are considered and tested against the design criterion). This simple solution approach is chosen since the main focus of this work is the multiscale design process and not the technique employed for solving the unit cell optimization problem. Genetic algorithms are employed in Ref. [24] for optimizing unit cells across significantly wide design spaces (i.e., large values of $l$). A heuristic technique is particularly suitable because the design space is typically highly multimodal and a concurrent evolution of multiple designs increases the probability of converging to near-global optimality [24].

### 3.3.2 Guidelines for Integration into Structure Design

Information passing between the materials and the structure design is critical in the MDD methodology. When there are size constraints on the structure, the design process needs to be integrated. Two different approaches are given below for integrating the two interrelated design domains.

*Approach I*

In this approach, specific values of non-dimensional frequencies drive the material unit cell objective function given by Eq. (15). For the case of isolating a harmonic excitation load, the excitation frequency $\omega^*$ is specified in non-dimensional form:

$$\Omega^* = \omega^* d / \sqrt{E_m/\rho_m}. \tag{21}$$

For the case of isolating an excitation with a broad frequency content, then a frequency range should be computed in non-dimensional form using Eq. (21) for the limits of this range. To obtain a value (or a range) for $\Omega^*$, the following 4 steps are followed:

*Step 1:*

Determine regions in the structure that are to be composed of banded materials. The structure in the bottom right of Fig. 1, for example, has two banded material regions. At this point several options emerge, two of which are: (i) place the banded materials near the excitation source to isolate the incoming waves from the onset; (ii) place the materials near the locations where isolation is desired.

*Step 2:*

Choose constituent materials with contrasting properties (i.e., $E_f$, $\rho_f$, $E_m$, $\rho_m$) for each banded region. Within the constraints of material availability, desired load-bearing capacity, and other engineering considerations, it is recommended to choose a material pair with high property ratios for the Young's modulus and density since the stronger the contrast in properties, the higher the potential for strong attenuation.

*Step 3:*

Choose the number of cells, $N_{Cell}$, to be used for attenuating wave propagation in the banded material regions of the structure (i.e., regions to be composed of the designed cells). In certain cases, it is appropriate to allocate $2N_{Cell}$ cells to a banded material region, as is the case in the region employed for isolating the force $F_2$ in the example structure shown in the bottom right of Fig. 1. Here, $N_{Cell}$ cells are included along each direction from the loading point. There are both upper and lower bounds on the value of $N_{Cell}$. Imposing a lower limit on the length of a single layer (e.g., due to manufacturing constraints), places an upper limit on the total number of layers, $n_{max}^{tot}$, for a banded material region with a prescribed length. Since a maximum of $n_{max}$ layers are allowed per cell (which is one of the constraints in the unit cell optimization problem), the upper bound of $N_{Cell}$ is enforced such that, $N_{Cell} \leq \lfloor n_{max}^{tot} / n_{max} \rfloor$, where the operator $\lfloor \gamma \rfloor$ is defined as the largest integer less than or equal to $\gamma$. On the other hand, a lower bound on $N_{Cell}$ is prescribed so that a sufficient attenuation (or isolation) is achieved. As discussed in Ref. [17], the lower bound on $N_{Cell}$ should not be less than 3.

*Step 4:*

Compute the actual cell length $d$ for each banded material region. This will depend on the lengths of the designated regions and the value of $N_{Cell}$ in each region. For example, if the structure is of length $L$, and if it is to be fully composed of a single type of periodic material, then $d = L/N_{Cell}$.

Upon evaluating the value (or the range of values) for $\Omega^*$, the objective function in Eq. (15) can be specified and the unit cell design process can be conducted. Once all the building block units cells are designed, the final structure design can be realized from the base homogenous material as well as the designed banded materials (synthesized cells).

*Approach II*

An alternative design approach involves first optimizing the unit cell (for a given banded material region and chosen material property ratios) for a maximum ratio of band-gap width to its central frequency, $\Delta\Omega^{n_{SB}} / \Omega_C^{n_{SB}}$, where $n_{SB}$, $\Delta\Omega^{n_{SB}}$ and $\Omega_C^{n_{SB}}$ denote band gap number, width and central frequency, respectively. The next step involves choosing $N_{Cell}$ and $d$ for each designated banded material region such that $\Omega_C^{n_{SB}} = \omega^* d / \sqrt{E_m / \rho_m}$ is satisfied for the known excitation frequency, $\omega^*$. For the broad-band case, a summation of the ratio of band-gap width to its central frequency over several band gaps is maximized, and appropriate scaling is followed based on the limits of the frequency range of interest.

*Remarks:*

(i) In Approach I the cell layout in the structure imposes a constraint on the material design through the value of the cell lengths $d$.

(ii) In Approach II, the material design fully precedes the structural design and hence imposes the cell length constraints (in conjunction with the property of the base material) on the structure design problem.

(iii) In all cases, the available space for the banded materials regions, the value of the excitation frequency, $\omega^*$, and the choice of material ratios ($\rho_f / \rho_m$ and $E_f / E_m$) collectively provide limiting conditions for the structure design. For example, as $\omega^*$ is decreased, the minimum required length of the relevant banded material region is increased (for a given choice of material property ratios).

## 4. Application of the MDD Methodology and Design Examples

In this section, the MDD methodology is demonstrated via numerical examples. Approach I presented in Section 3.3.2 is followed in all these examples. Approach II could have alternatively been followed without compromising the intended demonstration. It is assumed without loss of generality that there is no strict constraint for the total length $L$ of all the structures considered, and therefore in this section $d = 1$ for all the cells within the designed structures. In practice, the cell lengths are scaled according to the structure size constraints as discussed above, and consequently Steps 1 through 4 in Section 3.3.2 need to be followed. A direct scaling using Eq. (21) allows generalization to cases where there are size constraints at the structural level and hence on the lengths of the various banded material regions and the individual cells.

### 4.1 Isolation of Vibration: Attenuation of Distinct Frequencies

In this class of applications, the goal is to isolate the transmission of vibrations induced by an excitation at a single frequency and single position or by excitations at different frequencies and different positions. For the first case, a structure that is subjected to an excitation at one end and is fixed at the other end is considered. The implementation details of the material and structure design problems are presented in Sections 4.1.1 and 4.1.2, respectively. A case study demonstrating multiple excitations and reduced use of periodic materials is given in Section 4.1.3.

#### 4.1.1 Design of Material

In line with the proposed methodology, a unit cell is first designed to attenuate energy at the specified frequency of excitation. This material design requirement translates to a process of optimizing the unit cell layout (i.e., number of contrasting layers, and their relative thicknesses) in such a way as to have a frequency spectrum with a maximum attenuation at the frequency of interest, that is,

$$\text{Minimize:} \quad g = \xi_{\text{imag}}(\Omega^*), \tag{22}$$

where $\xi_{\text{imag}} = \text{Im}(\xi)$ and $\xi$ is a non-dimensional wavenumber defined as

$$\xi = k \times d. \tag{23}$$

The more negative this quantity, the stronger the attenuation [17], and therefore the more effective the vibration isolation.

Assume the available materials have property ratios of $\rho_f/\rho_m = 3$ and $E_f/E_m = 12$. Furthermore, assume that the layers are available in discrete values of thickness, e.g., $d^{(j)}/d = 0.1, 0.2, \ldots, 1$. To match up with these available layer thicknesses, $l = 10$, which in turn implies a search space spanning a total of $2^{10}$ alternatives.

For a particular excitation frequency, $\omega^*$, the above information is used in Eq. (21) to obtain the value of the target excitation frequency $\Omega^*$. In this case study, $\Omega^* = 20$. The optimal unit cell design was obtained following the approach described in Section 3.3.1. Fig. 2 shows the optimal design with encoding '0101010101' (in the inset) and its corresponding frequency spectrum. A wide stop band ($\Delta\Omega = 17.65$) centered at the target frequency is clearly observed. Noticeably the design consists of several sub-cells. Technically, the design could be represented by only one of the sub-cells. However, for maximum band-gap attenuation to match an already specified target non-dimensional frequency, five of these sub-cells are required to form the optimal unit cell. In other words, sub-cells are allowed to exist due to the inherent length scale in the design objective and are a natural consequence of the design problem.

**4.1.2 Design of Structure**

In this section, the synthesized unit cell presented in Section 4.1.1 is used to form a structure that isolates a harmonic excitation load applied at one end from being transmitted to the opposite end. Two structure design approaches are presented: (i) a simple approach in which the entire structure is "filled" with the designed cell in a manner that does not allow for using partial cells or leaving any space not formed from the cells (yielding a fully periodic structure), and (ii) a more advanced approach that allows for a homogenous region in the structure to be included to alter the transmission at the receiving end (yielding a partially periodic structure). These alternative approaches are discussed next.

### 4.1.2.1 Fully Periodic Structure

In this approach, the structure is completely formed from the optimal unit cell design. The structure has length $L = 5$ and is formed of five cells, as shown in Fig. 3. This structure is "fixed" to a rigid wall at the right end (zero prescribed displacement), and is subjected to a harmonic excitation in the form of prescribed displacement applied at the left, that is, $u(0,t) = \sin(\omega^* t)$. The frequency of excitation is $\omega^* = \Omega^* \sqrt{E_m/\rho_m}/d$, where $\Omega^* = 20$ (the frequency that the unit cell is designed to attenuate), $d = 1$, and $\rho_m$ and $E_m$ are respectively chosen in a manner consistent with the ratios $\rho_f/\rho_m = 3$ and $E_f/E_m = 12$. Supplemented by the above boundary conditions and with quiescent initial conditions, Eq. (1) was solved using a finite element model consisting of 600 piecewise linear elements. The finite element mesh was chosen to satisfy the condition $h_f^e/c_f = h_m^e/c_m$, where $h_f^e$ and $h_m^e$ denote the size of a fiber element and matrix element, respectively. Time integration was carried out using a standard explicit central difference scheme with a time step of $\Delta t = h_f^e/c_f = h_m^e/c_m = 3.125 \times 10^{-3}$ seconds. Fig. 4 shows the maximum point-wise spatial response (spatial envelope) in the entire structure computed over a time span of 12 seconds. As a measure of transmission, the maximum displacement within the 5th cell (see Fig. 3) is observed to have a maximum value of approximately 52% of the input displacement amplitude during the 12-second simulation. Clearly, a significant degree of vibration isolation is achieved, especially considering that no dissipative mechanisms are involved. Should damping be included, additional attenuation would be expected.

A thorough discussion of the observed dynamical behavior has been provided in an earlier paper [17]. In summary, due to the discrete constituent material mismatch in the cells, wave scattering takes place and, therefore, at any given time there are both forward and backward traveling waves in the structure. The speed of wave propagation in each material layer depends on the elasticity modulus and the density of the material. Consequently, in-phase and out-of-phase forward and backward traveling waves meet at rates that are dictated by the topology of the cell and the temporal frequency of wave motion. Depending on the dispersion, either the in-phase or the out-of-phase interferences dominate. At pass-band frequencies, the former dominates and effective wave propagation is enhanced. Conversely, at stop-band frequencies, the latter is dominant and overall wave propagation is inhibited. In the latter case, the displacement is attenuated spatially with an increase in the distance from the excitation point. From the perspective of energetics, a low amount of energy is pumped into the structure at stop-band frequencies (compared to pass-band frequencies) due to the overall low response throughout the wave field.

*Remarks on Layer Ordering:*

Note that the first layer in each cell in Fig. 3 is chosen to be of the compliant type so that the last one is of the stiff type. As far as correlation between the infinite material and finite structure is concerned, the layer ordering is irrelevant due to symmetry. However, for minimizing vibration transmission the ordering is important. This is because displacements within stiff layers are relatively less than in adjacent compliant layers for a given internal force. Hence, when the last layer is chosen to be stiff, the motion close to the right end is further decreased, and in turn, the force transmission at that end is further reduced (compared to having the reverse ordering of the layers). To quantify this difference, the traction at the wall was computed for both cases of material ordering. Traction values of $\sigma_t^{max} = 53$ and $\sigma_t^{max} = 530$ were obtained when the last (rightmost) element was fiber (stiff) and matrix (compliant), respectively.

**4.1.2.2 Employment of "Tuning Layer"**

If a small additional homogenous layer is added to the synthesized bounded structure, a change in the overall "wave phasing" takes place and causes either an improvement or a deterioration of wave attenuation in the structure as a whole. This added layer is referred to as a *tuning layer* and its length is denoted $d_{TL}$. To study its effect, the problem considered in the previous section (Section 4.1.2.1) was repeated with the addition of a tuning layer; see Fig. 5 for a schematic. Each of the two material types was employed in the tuning layer. In addition, the cell layering shown in Fig. 3 was considered along with the reverse ordering (encoded '1010101010'). Fig. 6 shows the computed traction at the rigid wall (for the same 12 seconds time period) as a function of $d_{TL}$, cell layer ordering and tuning layer material choice. When the tuning layer is composed of the stiff material, the transmitted stress is smaller than the compliant material case. Furthermore, regardless of which layer ordering is used, there is an optimal tuning layer length for which the transmitted traction is minimized. This optimal value repeats with $d_{TL}$ in a nearly periodic fashion, which is a consequence of the wave phasing phenomenon. High tuning layer sensitivity is apparent for the two designs in which the materials of the tuning layer and the last (rightmost) layer of the rightmost cell are not the same. The best design is '1010101010/Stiff' as it exhibits small transmitted stress and is not sensitive to $d_{TL}$.

Incorporating a tuning layer requires modifying Step 4 of Approach I for structure design (see Section 3.3) as follows:

*Approach I - Step 4 ( for Structure with Tuning Layer)*

Determine the type (recommended to be stiff) and size, $d_{TL}$, of the tuning layer. Considering that the designated banded material region (or full structure as in example) minus the tuning layer would be composed of $N_{Cell}$ cells of a single design, then the actual cell length is to be computed using the formula $d = (L - d_{TL})/ N_{Cell}$. For the practical case in which $L$ is a given constraint, the complete design process should be repeated for different values of $d_{TL}$ and the design corresponding to the best response should be selected.

### 4.1.3 Multiple Excitations and Reduced Use of Banded Materials

In this section, multiple excitations (at different frequencies and positions) are considered. Since only a small number of cells need to be employed for matching the periodic material behavior, a reduced number of optimally synthesized banded materials can be used to form the bounded structure. Consider a structure that is fixed at the right end and has length $L = 21$. This structure is subjected to two harmonic forcing loads, one at $x = 0$ and one at $x = 12$. Furthermore, let each of the loads have a different frequency, e.g., $f_1(t) = f(0,t) = F_1 e^{i\omega_1^* t}$, $-\infty < t < \infty$ and $f_2(t) = f(12,t) = F_2 e^{i\omega_2^* t}$, $-\infty < t < \infty$, where $\omega_1^*$ and $\omega_2^*$ are chosen such that $\Omega_1^* = 20$ and $\Omega_2^* = 17.5$, respectively. The force amplitudes are chosen to be equal, i.e., $F = F_1 = F_2$, and $F$ is chosen arbitrarily since the reported displacements are non-dimensionalized with respect to the static deflection of the left end ($F_1 = F$, $F_2 = 0$), $\delta = FLd/E_{\text{avg}}$, where

$$E_{\text{avg}} = \left( r_{\text{f}} \frac{1}{E_{\text{f}}} + r_{\text{m}} \frac{1}{E_{\text{m}}} \right)^{-1}, \qquad (24)$$

and where $r_{\text{f}}$ and $r_{\text{m}}$ denote the volume fraction of each of the constituent materials in the structure as a whole.

Using the procedure described in Section 3.3.1, and using Eq. (22) for the objective function, two unit cell designs are sought, one for attenuating harmonic response at $\Omega^* = 20$ and one for attenuating harmonic response at $\Omega^* = 17.5$. The first design was presented in Section 4.1.1, with encoding '0101010101'. The optimized unit cell design for attenuating harmonic response at $\Omega^* = 17.5$ is encoded '1101011010' and is shown (in the inset) along with its frequency band structure in Fig. 7.

In forming the bounded structure, 3 cells of the first design are located at the left end (solely to attenuate the response resulting from $f_1$), and 6 cells of the second design are centered around $x = 12$ (solely to attenuate the response resulting from $f_2$). A schematic of the designed partially periodic structure is given in Fig. 8a. The finite element model used consists of 1992 piecewise linear elements with lengths again satisfying the condition $h_{\text{f}}^{\text{e}} / c_{\text{f}} = h_{\text{m}}^{\text{e}} / c_{\text{m}}$. To test the performance of this configuration, the normalized maximum forced steady-state response in the entire structure, $\max_{0 \leq x \leq L}(U/\delta)$, was computed over the excitation frequency range $0 \leq \Omega^* \leq 50$ (with steps in frequency of 0.25) for the following two cases: (i) $f_1 \neq 0$, $f_2 = 0$, and (ii) $f_1 = 0$, $f_2 \neq 0$. These results are shown in Figs. 9a and 9b, respectively. The response corresponding to a statically equivalent homogenous structure (see Fig. 8b) modeled with 1680 finite elements is overlaid for comparison. The properties of this equivalent structure are $E_{\text{avg}}$ and $\rho_{\text{avg}}$, where the latter denotes the averaged density:

$$\rho_{\text{avg}} = \frac{1}{d}(d_f \rho_f + d_m f_m). \qquad (25)$$

It is observed from Fig. 9 that the maximum response in the entire structure is reduced significantly in the stop-band frequency ranges corresponding to each of the designed materials (unit cell designs) employed when only the excitation at the respective frequency is non-zero. The steady-state spatial response of the engineered banded structure subjected to both excitations simultaneously is shown in Fig. 10, demonstrating the ability of the banded materials (groups of synthesized cells) to confine motion to the excitation locations. Fig. 10 also shows a significantly reduced response at the excitation locations compared to the response of the equivalent homogenous structure.

This case study reveals that (i) the global dynamical resonance response of a bounded structure could be made to conform to the local dynamical properties of a banded material used in only a small portion of the structure, and (ii) this congruence in response could be simultaneously realized for more than one localized banded material distributed in different regions of the structure. However, these outcomes are not guaranteed unconditionally. Due to the aperiodicity of the structure as a whole, it is possible for global resonant modes to undesirably appear within stop-band frequencies of the banded materials. From the practical perspective, the designer can experiment with several configurations (varying the number, the layer ordering, and possibly the designs, of the employed cells) until a desired outcome is realized. The authors' experience with other case studies suggest that it is not difficult to achieve effective and reduced employment of banded materials for vibration isolation for cases involving multiple excitations.

**4.2 Isolation of Shock: Attenuation of a Wide Range of Frequencies**
**4.2.1 Design of Material**

In this section, a unit cell is designed to attenuate wave propagation over a significantly wide frequency range. The motivation is to use the designed periodic material for shock isolation. A shock load corresponds to a transient pulse that typically has broad frequency content. It may be impossible to synthesize a stop band with enough bandwidth to cover the whole frequency range of the pulse. Therefore, the goal is to tailor a unit cell that maximizes the total sum of stop-band frequency ranges. The corresponding design objective then becomes minimizing the transmissibility, $T_\text{p}$ (given in percent), within the frequency range of interest, that is,

$$\text{Minimize:} \quad g = T_\text{p} = 100 \times \frac{\int_{\Omega_\text{min}}^{\Omega_\text{max}} H(\xi_\text{real}(\Omega)) \text{d}\Omega}{\Omega_\text{max} - \Omega_\text{min}}, \tag{26}$$

where $H(\xi_\text{real}(\Omega))$ is a hard limit function defined as

$$H(\xi_\text{real}(\Omega)) = \begin{cases} 1 & \text{if } \xi_\text{real} > 0 \text{ (pass band)} \\ 0 & \text{if } \xi_\text{real} = 0 \text{ (stop band)} \end{cases}, \tag{27}$$

and where the frequency range of interest is [$\Omega_\text{min}$, $\Omega_\text{max}$].

This design problem was implemented for a shock load with frequency content $0 \leq \Omega^* \leq 50$, subject to the same geometric and layer thickness availability constraints for the material design problem of Section 4.1.1. The optimal design and the corresponding band diagram are shown in Fig. 11. The optimal unit cell here has the encoding '0010110111'. The achieved value of transmissibility is $T_\text{p}$ = 16.3%, compared to $T_\text{p}$ = 33.2% for the unit cell design of Fig. 1. For a homogenous medium, $T_\text{p}$ = 100%.

### 4.2.2 Design of Structure

Two approaches are presented for the structural design phase of the present problem: (i) a simple approach based on a single periodic material spanning the entire structure (fully periodic structure), and (ii) a more advanced approach that utilizes more than one periodic material design.

### 4.2.2.1 Fully Periodic Structure

Fig. 12 depicts the fully periodic structure comprising five cells of the material designed for shock isolation (encoded '0010110111'). The shock load is represented by a prescribed displacement at the left end having the form of a double-Gaussian function:

$$u(0,t) = e^{-a(t-b)^2} - e^{-a(t-c)^2}, \quad (28)$$

where *a*, *b* and *c* are parameters. To synthesize a signal with a frequency content that approximately spans the range of interest, $0 \leq \Omega^* \leq 50$, the parameters were chosen as:

$$a = 450, b = 0.25, c = 0.26. \quad (29)$$

The spatial domain is discretized into 620 piecewise linear finite elements (chosen such that $h_f^e / c_f = h_m^e / c_m$ is satisfied), and a time step of $\Delta t = h_f^e / c_f = h_m^e / c_m = 3.1 \times 10^{-3}$ is used for the numerical integration.

The computed displacement time histories for the left end (input) and point O (representing an output position) are shown in Fig. 13. The maximum displacement at point O is approximately 7% of the maximum input displacement. The maximum displacement within the 5th cell (see Fig. 12) is approximately 45% of the maximum input displacement during the 12-second simulation, as shown in Fig. 13. The spatial envelope of the maximum displacement response for the entire structure is presented in Fig. 14. It is apparent that the transmission of the shock is effectively isolated within the designed banded structure.

### 4.2.2.2 Stacking of Multiple Cell Designs for Enhanced Isolation

Additional attenuation strength beyond that obtained by the structure of Fig. 12 can be achieved by several means: (i) expanding the design space of the unit cell to include a wider range of available layer thicknesses, (ii) including more cells in the finite structure (the effect of the number of cells is discussed in Ref. [17]), (iii) stacking groups of cells of multiple designs, and/or (iv) intentionally introducing disorder (in geometry or material properties) to the cell layers in a manner that would enhance attenuation (see Section 5). The third route is pursued in the present section.

Multiple cell designs can be generated such that the stacking (in groups) of the different cell designs produces a lower transmissibility that might be possible with a single optimized cell design. This effect, however, will take place over a larger region in the structure than when a single material design is employed. To demonstrate this process, the optimal design, encoded '001010111', was chosen for the first 5 cells of a 10-cell long structure. A different design was generated for the remaining 5 cells with the goal to fill in with stop bands the pass-band frequency ranges of the optimal design. The best design in satisfying this goal is encoded '1111000010'; the unit cell and corresponding frequency spectrum are shown in Fig. 15. A comparison of the band structure layout for both designs is illustrated in Fig. 16. In the figure, a vertical frequency scale is presented, and relative to this scale the computed stop-band frequency ranges of each design are shaded in grey (in the first two columns). The third column was generated by calculating the union of the stop-band frequency ranges for the two designs. The term "combination" will be used to refer to this resultant frequency band layout. For the secondary design encoded '1111000010', $T_p$ = 20.3%, while for the two designs in "combination", $T_p^{comb}$ = 7.8%. Hence by stacking multiple designs, the overall "non-local" material transmissibility is reduced by approximately 52% (compared to the optimal design considered alone).

To investigate the actual effects on a finite structure, the problem shown in Fig. 17 was considered ($L$ =10). The total time duration of the simulation was extended to 18 seconds to account for the larger structure length (compared to the previous example). A total of 1240 piecewise linear finite elements were used for the spatial discretization (following $h_f^e / c_f = h_m^e / c_m$), and a time step of $\Delta t = h_f^e / c_f = h_m^e / c_m = 3.1 \times 10^{-3}$ was used for the temporal integration. The spatial envelope of the maximum displacement response for the entire structure over a simulation duration spanning $0 \leq t \leq 18$ is presented in Fig. 18, indicating very low response in the region close to the right end. The highest values of maximum displacement within the 5th and 10th cells are approximately 37% and 5%, respectively, of the maximum input displacement. For comparison, the problem was solved employing 10 cells with encoding '001010111'. The computed maximum displacement in the 10th cell of this single cell-type structure was 33% of the maximum input displacement, compared to 5% for the multiple cell-type structure. Further insight can be gained by studying time histories and corresponding frequency spectra of traction, $\sigma_t$, at the right end for both the multiple cell-type structure and the single cell-type structure. These results are shown in Fig. 19. The frequency spectra are obtained by a standard Fast Fourier Transform algorithm. The traction at the right end of the multiple cell-type structure has a maximum magnitude that is approximately 27% lower than that for the single cell-type structure. The frequency spectra show that the spread of energy at the right end of the multiple cell-type structure is consistent with the frequency bands of the two designs in "combination", and the frequency spread of the single cell-type structure is consistent with the frequency bands of the '0101010101' periodic material.

The results demonstrate that there is a substantial improvement in shock isolation within a structure composed of multiple cell-types (chosen in the manner discussed above) compared to those for a structure composed of only a single optimally designed unit cell type. The improvement at the structural level is even higher than the improvement (decrease in $T_p$) at the unit cells level (when comparing $T_p^{comb}$ of the "combined" set to the $T_p$ of the optimal design encoded '0101010101'). This example indicates that while a material toplogy could be locally optimal, it might not be so at the global level. This insight is central to the multiscale dispersive design paradigm.

**4.3 Control of Energy Propagation Speed**

The proposed MDD methodology can be used to regulate the speed of dominant wave propagation, which is known to correspond to the speed of energy propagation [1]. In Ref. [17], two distinctive wave fronts were observed in a structure. The faster and slower wave velocities were approximately equal to the phase and group velocity values (dispersion curves quantities for corresponding periodic medium) within the frequency range of the most excited pass band. Consequently, cells can be designed to have band diagrams tailored for particular phase and group wave velocities, and subsequently can be used to form structures that allow approximately the same speeds of wave propagation.

A unit cell encoded '0000110000' (previously studied in Ref. [17]) was used to form a finite structure as shown in Fig. 20. The structure consists of 5 cells (each of length $d = 1$) stacked at the left end. A long homogenous portion at the right end is included so that the end point has a non-zero displacement that could represent the "output" of the periodic portion of the structure. A prescribed displacement is applied at the left end at a frequency of $\Omega^* = 17.5$. A total of 650 piecewise linear finite elements were used for the spatial discretization (following $h_f^e/c_f = h_m^e/c_m$), and a time step of $\Delta t = h_f^e/c_f = h_m^e/c_m = 0.01$ was used for the temporal integration. Ten different cases were considered with varying material properties. The ratio of phase velocities was kept constant at $c_f/c_m = 2$, while the ratio of densities, $\rho_f/\rho_m$, was varied in integer increments between 1 and 10. In all cases, the frequency content of the propagating waves fell mostly within the unit cell's 2nd pass band (see Fig. 21 for three representative cases). Based on this observation, the maximum group velocity, $\max(c_g)$, at the 2nd pass band was calculated for each case. A time-space displacement contour plot was then generated for each case, and dominant, observable wave speeds, $\alpha$'s, were calculated (see Ref. [17] for details). The observable wave speeds are compared to the infinite periodic material group velocities in Fig. 22. It is observed that the infinite and finite velocities are very close. The stop-band/pass-band map for the range of material properties considered is shown in Fig. 23. The decrease in the group velocity with increase in density ratio, as presented in Fig. 22, is explained by the narrowing of the 2nd pass band with increasing density ratio, as shown in Fig. 23. From the dispersion curves, it can be observed that a narrow pass band has a low maximum group velocity.

These results reinforce the analysis conducted in Ref. [17] and demonstrate how the building block unit cell could be designed for a desired speed of energy propagation (that corresponds to the group velocity in the constituent periodic medium) in the finite periodic structure.

## 5. Geometric Disorder Sensitivity

An important practical consideration in the design of banded materials and structures is the sensitivity of their dynamical characteristics to disorder (manifested, in 1D, as variations in the layer thicknesses or the layer material properties[2]). Several studies have been done on the effects of disorder on band-gap size and density of states (DOS) of photonic crystals as well as on the transmission in these crystals (see, for example, Refs. [25-27]). A common approach is to introduce a random disorder on the location or geometric features of the inclusions (such as cylinders or spheres) across several cells, and to subsequently solve for the band structure, DOS, or transmission performance, of a "supercell" that comprises the perturbed cells. In the present study, disorder is introduced to a unit cell (through perturbation of the thickness of each of the layers), and its effects on the band structure attenuation characteristics, including transmissibility, are analyzed as a function of disorder strength. It suffices to study the periodic material only, since the outcome is expected to correlate with that of a corresponding finite structure.

The optimal periodic materials '0101010101' (shown in inset of Fig. 2) and '0010110111' (shown in inset of Fig. 11) were considered. The attenuation-related performance measures are given in Eqs. (22) and (26), respectively. The sensitivity of these performance quantities to small random variations in the layer thicknesses was investigated – at the target frequency of $\Omega^* = 20$ for the first unit cell, and across the range of $0 \leq \Omega^* \leq 50$ for the second.

A random number generator was used to realize 1000 disorder patterns comprising different cell layer thicknesses, obtained from a uniform distribution with zero mean and varying values of standard deviation (0.5% to 5% in increments of 0.5%). Let the percentage perturbation of the thickness of layer $j$ within the unit cell be denoted $\delta^{(j)}$. For a given $\delta$-distribution characterized by percentage standard deviation $\sigma_\delta$, the upper limit percentage value is given by

$$\delta_{max} = \sqrt{3}\sigma_\delta. \tag{30}$$

---

[2] As shown in eq. (7), a variation in either the layer thickness or the material properties results in a variation in the product $k^{(j)}d^{(j)}$. A variation in the material properties also affects $Z^{(j)}$.

Denoting $r^{(j)}$ for a random number associated with layer $j$ and spanning the range $-1 \leq r^{(j)} \leq 1$, $\delta^{(j)}$ is given by

$$\delta^{(j)} = \frac{\delta_{max}}{100} r^{(j)}, \tag{31}$$

that is,

$$-\frac{\delta_{max}}{100} \leq \delta^{(j)} \leq \frac{\delta_{max}}{100}. \tag{32}$$

Therefore, the actual statistically perturbed thickness of layer $j$ is given by

$$d_\delta^{(j)} = d^{(j)} + \delta^{(j)} d^{(j)}, \tag{33}$$

implying that the total length of the cell,

$$d_\delta = \sum_{j=1}^{N_{cell}} d_\delta^{(j)}, \tag{34}$$

is now perturbed from unity.

Upon running a Monte Carlo simulation based on the above perturbations, frequency of occurrence of variations of the performance quantity was computed, for the various values of $\sigma_\delta$. Fig. 24 presents sensitivity analysis information for the unit cell '0101010101' designed for vibration isolation. The abscissa defines the change of the imaginary wavenumber, $\Delta\xi_{min}$, at the frequency of interest $\Omega^* = 20$, that is, at the center of the wide stop band. A positive change means a loss in attenuation strength. The ordinate defines frequency of occurrence as a percentage value. Fig. 25 shows the same results in the form of percentiles. For a 5% degree of disorder in the geometric features of the unit cell, the negative deterioration of performance does not exceeds 4%. Note that 95% of the population of perturbed unit cells do not suffer from performance deterioration exceeding approximately 2%. Interestingly, the results also show that disorder could play a positive role and improve attenuation by up to approximately 2.5%. The process of "intentional disordering" for improvement of performance is not pursued in this study. Fig. 26 presents sensitivity analysis information for the unit cell '0010110111' designed for shock isolation. Here the abscissa defines the change of transmissibility, $\Delta T_p$, within the frequency range of interest $0 \leq \Omega^* \leq 50$. As for the harmonic case, a positive change means a loss in attenuation strength. Fig. 27 shows the same results in the form of percentiles. Geometric disorder of 5% results in transmissibility increases of less than 7%. The 95th percentile value of transmissibility increase is less than 2%. On the flip side, a decrease of transmissibility (i.e., increase in attenuation strength) of up to approximately 15% may be achieved due to disorder.

The results given in Figs. 24-27 indicate that optimal designs of periodic materials, and hence designs of periodic structures built from these materials, enjoy an excellent degree of robustness to geometric disorder.

## 6. Summary and Conclusions

The multiscale dispersive design methodology has been proposed for the passive non-dissipative control of elastic wave motion in heterogeneous structures. This novel hierarchical design process consists of two phases: (i) designing the periodic material configuration given by the unit cell(s), i.e., number of layers, their type, and their thicknesses, for desired dispersion characteristics (frequency band structure), and (ii) employing the designed cell(s) to form a fully periodic or partially periodic structure, to achieve the desired structural response objective. An integer programming formulation for the 1D unit cell design problem was given, and guidelines were presented for integrating the cell design into the structural design problem. The effectiveness of the proposed methodology was demonstrated by investigating case studies involving isolation of single harmonic and wide-frequency excitations as well as control of the speed of energy propagation. Various design scenarios incorporating the developed methodology were given, including the inclusion of a tuning layer, the design for multiple excitations, and the reduced use of designed periodic materials. The use of multiple cell designs (grouped in stacks) was proposed and shown to be a promising approach to engender broad frequeny bandwidth isolation. Statistical analyses revealed that the designed unit cells exhibit a high degree of robustness with respect to geometric variation, which suggests that the methodology is well suited for practical design applications.

The proposed MDD methodology can be applied directly to structures with other types of periodicities, such as structures exhibiting geometric periodicities or boundary periodicities. In the same manner, periodic unit cells can be designed for desired frequency characteristics and, with appropriate scaling, used as building blocks to form a global structure with related dynamical characteristics.

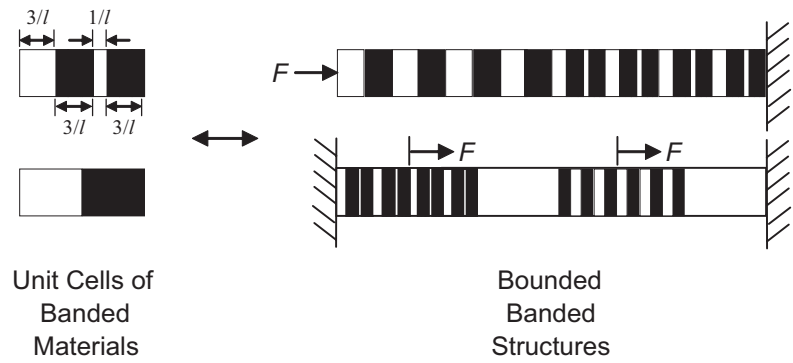

Figure 1  Schematic showing examples of synthesized banded materials (left) and examples of bounded structures composed of these materials (right).

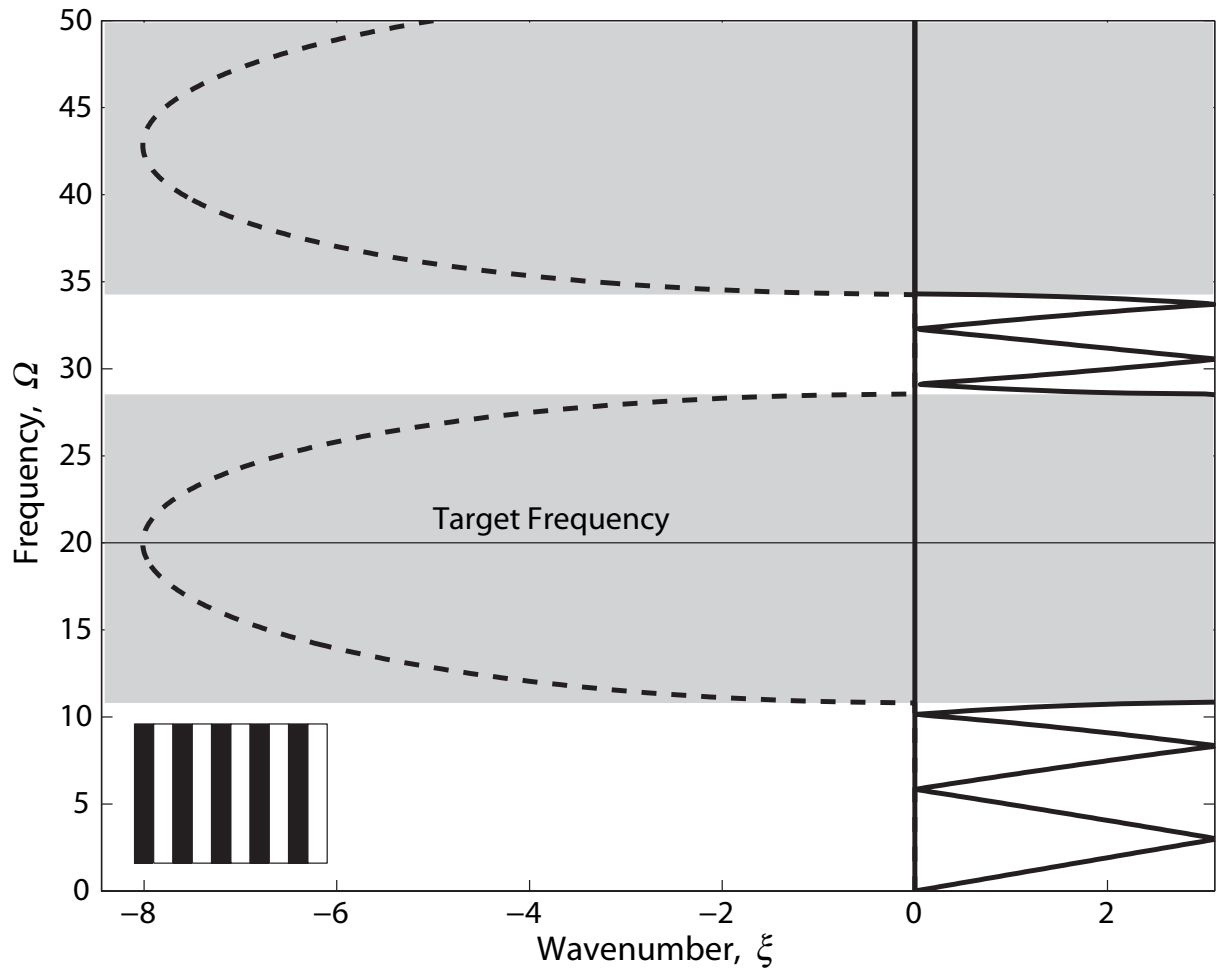

Figure 2  Frequency spectrum for optimal unit cell design for case involving harmonic excitation at non-dimensional frequency $\Omega^* = 20$. ( —— ) Real part (pass band), ( – – ) Imaginary part (stop band). Stop bands are shaded. Design is shown in inset and has a binary code of '0101010101' (black: stiff material, white: compliant material).

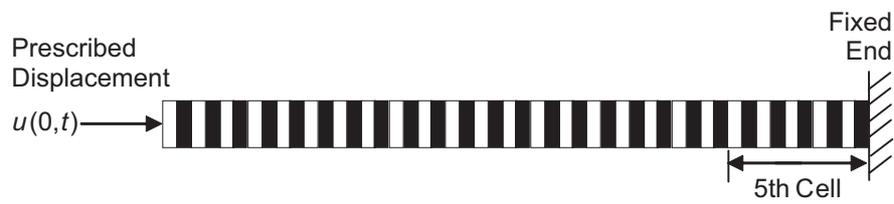

Figure 3  Banded structure designed for achieving vibration isolation. The structure is formed from the synthesized unit cell encoded '0101010101', shown in Fig. 2, with reverse order of layering to minimize displacement at the receiving end. Its length is $L = 5$. The excitation is harmonic at $\Omega^* = 20$ and is applied at the left end.

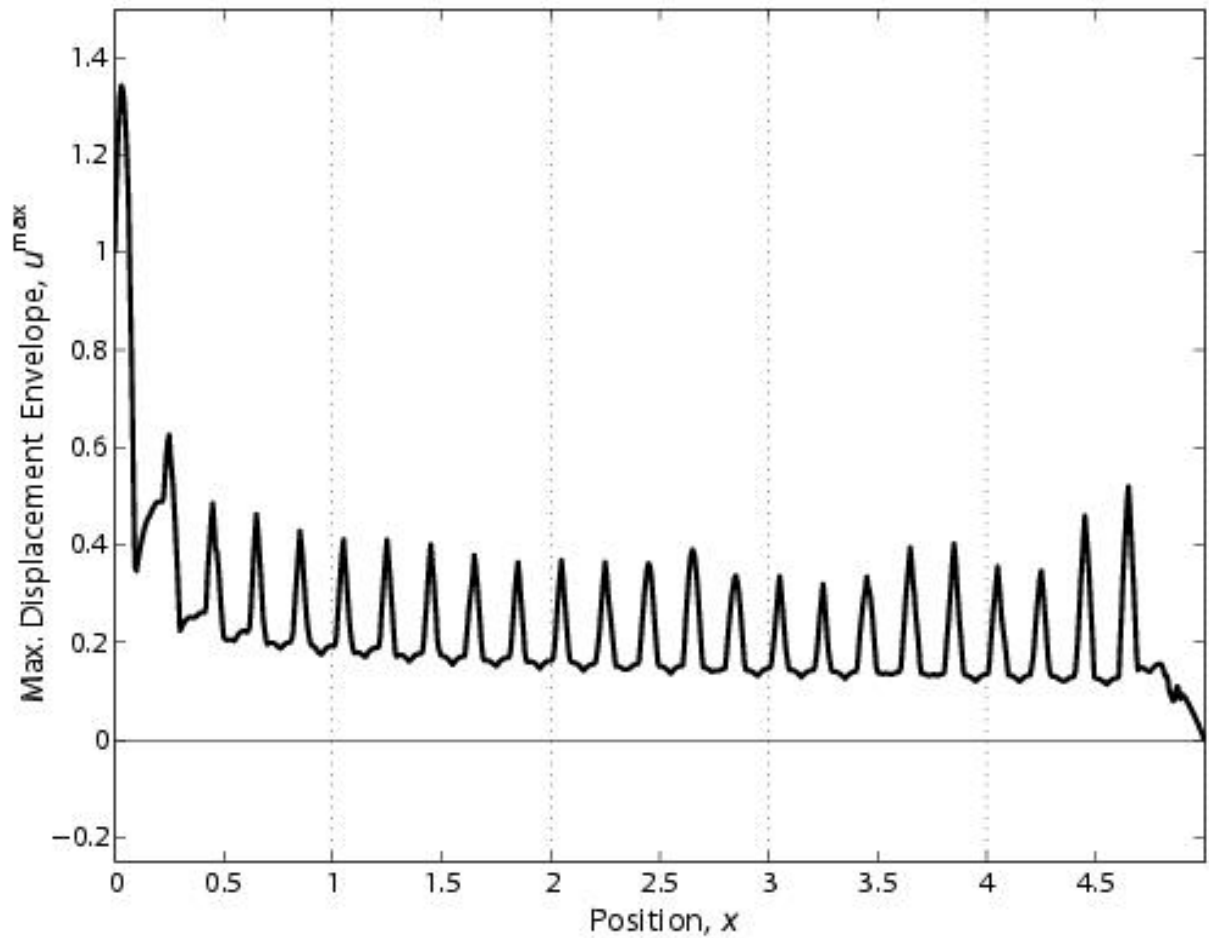

Figure 4　Envelope of maximum displacement throughout duration of numerical simulation (12 seconds) in banded structure designed for vibration isolation. The excitation is harmonic at $\Omega^* = 20$ and is applied at the left end of the structure.

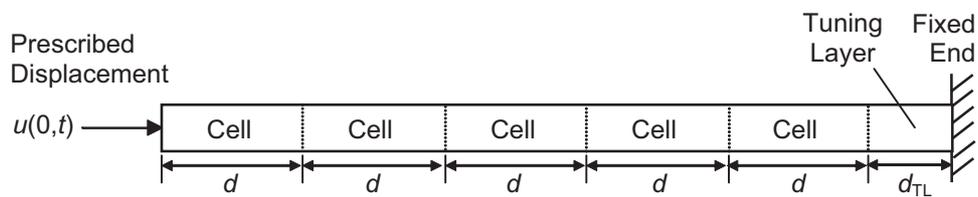

Figure 5   Schematic showing a synthesized structure composed of five cells and a tuning layer placed at the far right end. Length of tuning layer is denoted $d_{\text{TL}}$. The excitation is harmonic at $\Omega^* = 20$ and is applied at the left end.

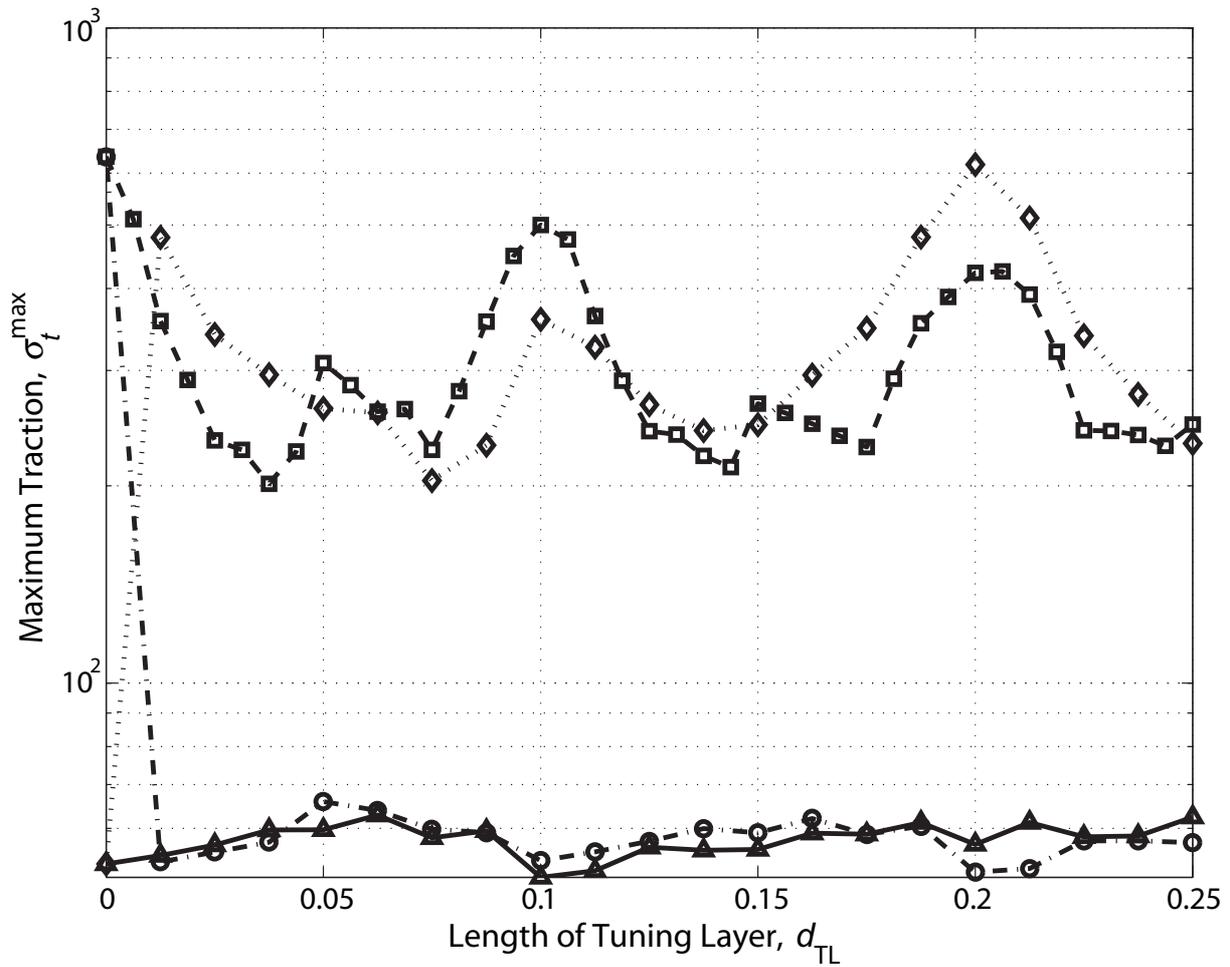

Figure 6  Maximum transmitted traction at right (receiving) end versus length of tuning layer, for different layer ordering and different tuning layer material type. A binary encoding scheme is used to represent the layered unit cell. Stiff and compliant tuning layers are denoted "/Stiff" and "/Compliant", respectively. (─○·) '0101010101'/Stiff, (─□─) '0101010101'/Compliant, (─△─) '1010101010'/Stiff, (··◇··) '1010101010'/Compliant.

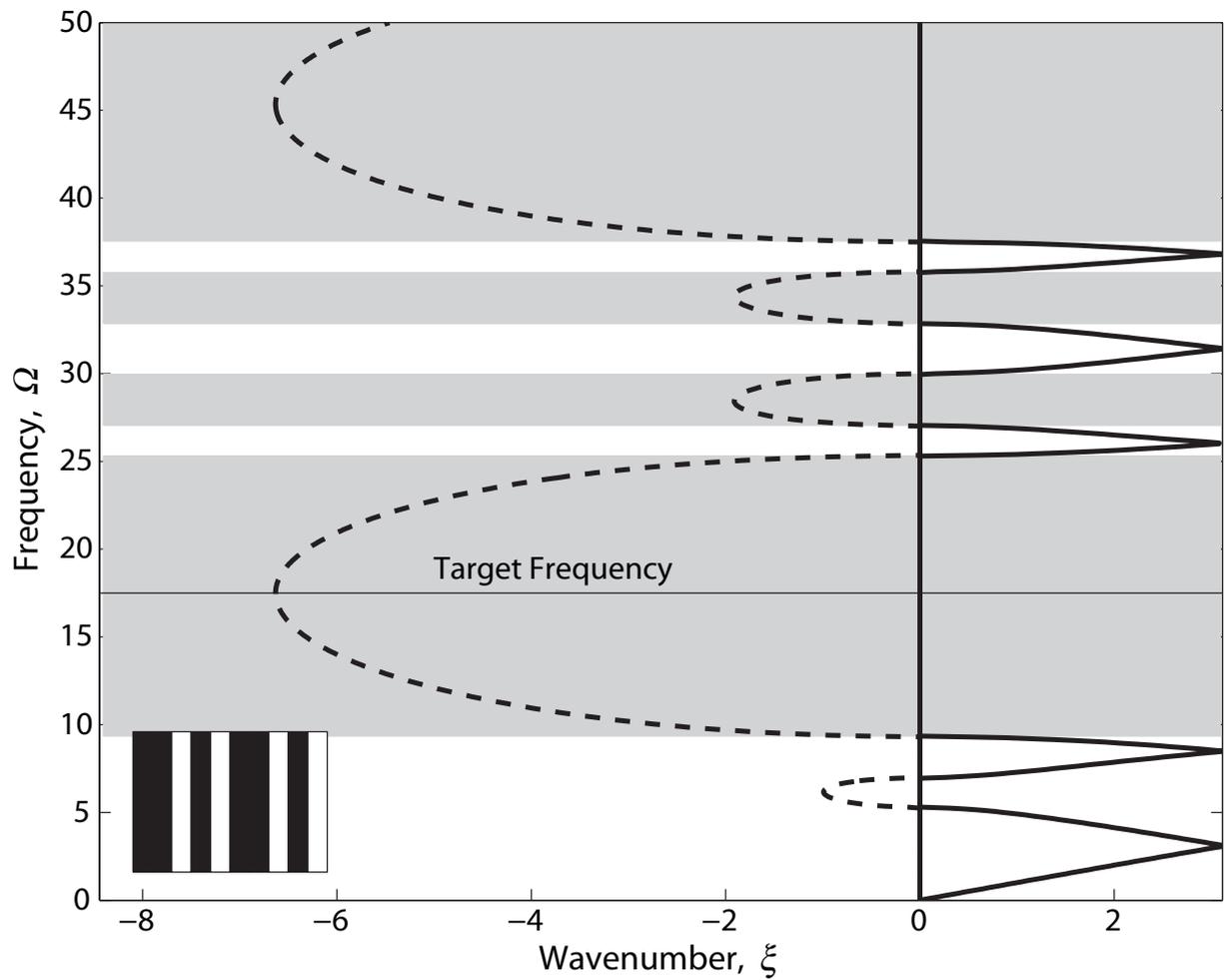

Figure 7  Frequency spectrum for unit cell design optimized for maximum attenuation at non-dimensional frequency $\Omega^* = 17.5$. (———) Real part (pass band), (— —) Imaginary part (stop band). Stop bands are shaded. Design is shown in inset and has a binary code of '0010100101' (black: stiff material, white: compliant material).

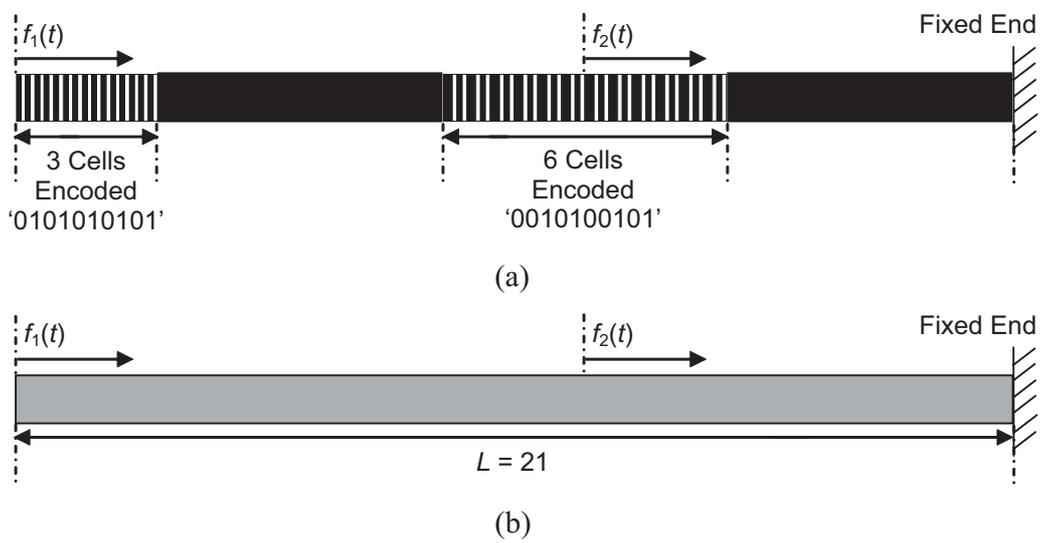

Figure 8 (a) Finite structure partially composed from cells encoded '1010101010' (Fig. 1) and '0010100101' (Fig. 7) while most of the structure is homogenous. (b) Statically equivalent homogenous structure. The length of both structures is $L = 21$. Excitation locations are shown.

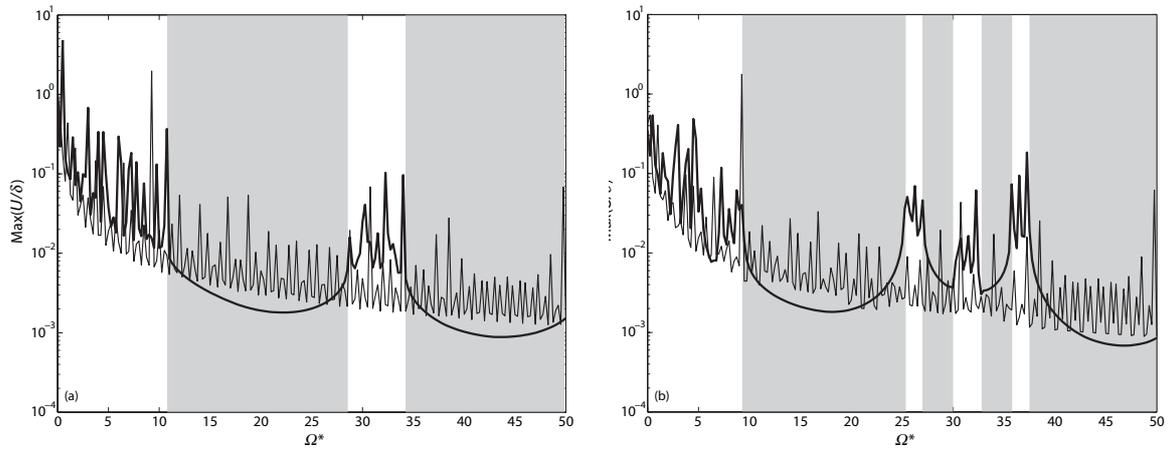

Figure 9  Frequency response of partially periodic structure shown in Fig. 8 when subjected (a) only to excitation $f_1$, and (b) only to excitation $f_2$. Stop-band frequency ranges are shaded for unit cell encoded (a) '1010101010', and (b) '0010100101'. The response of the equivalent homogenous structure is superimposed for both cases. (━━━) Banded structure, (───) Homogenous structure.

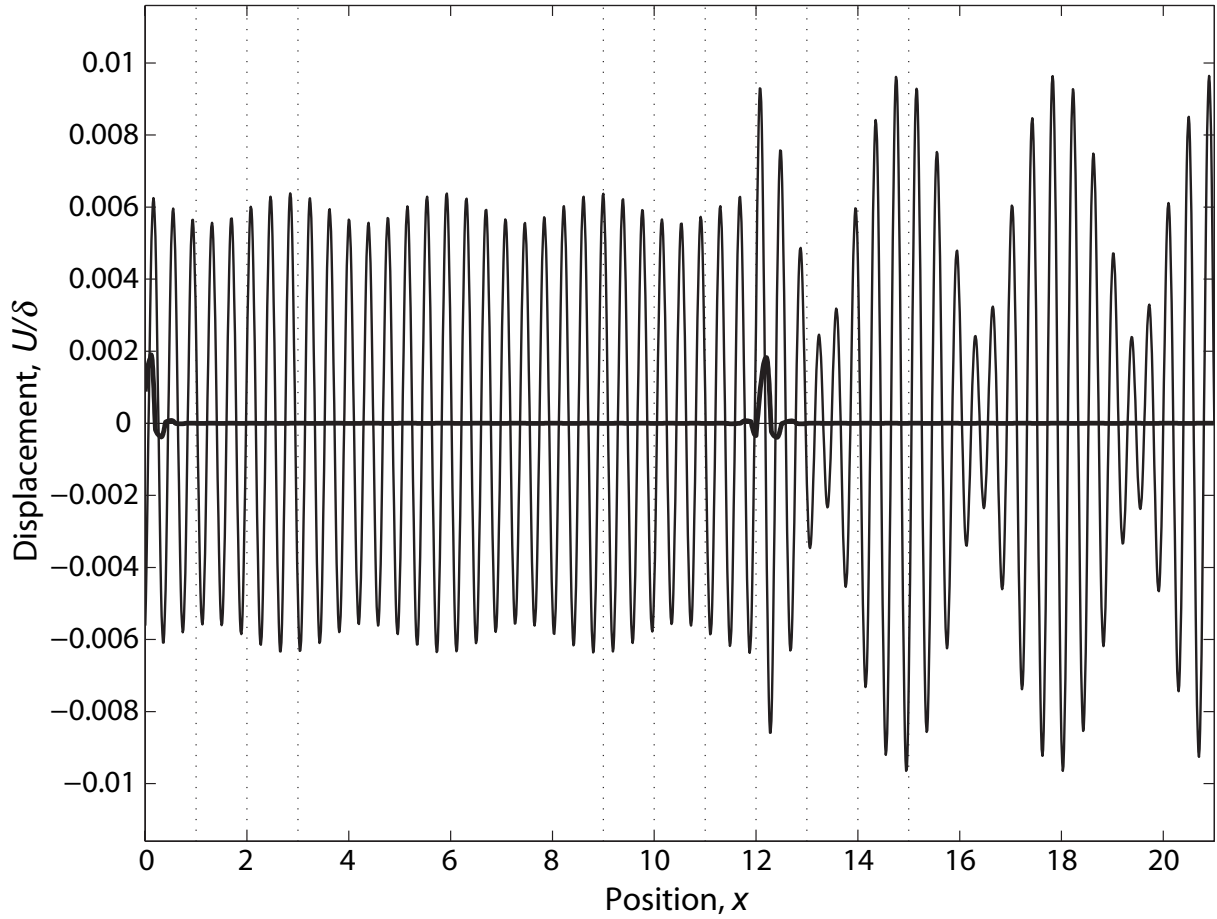

Figure 10   Spatial response of partially periodic structure shown in Fig. 8a when subjected to both excitations $f_1$ and $f_2$. The response of the equivalent homogenous structure is superimposed. (━━) Banded structure, (───) Homogenous structure. Vertical dotted lines indicate boundaries of designed periodic cells within structure.

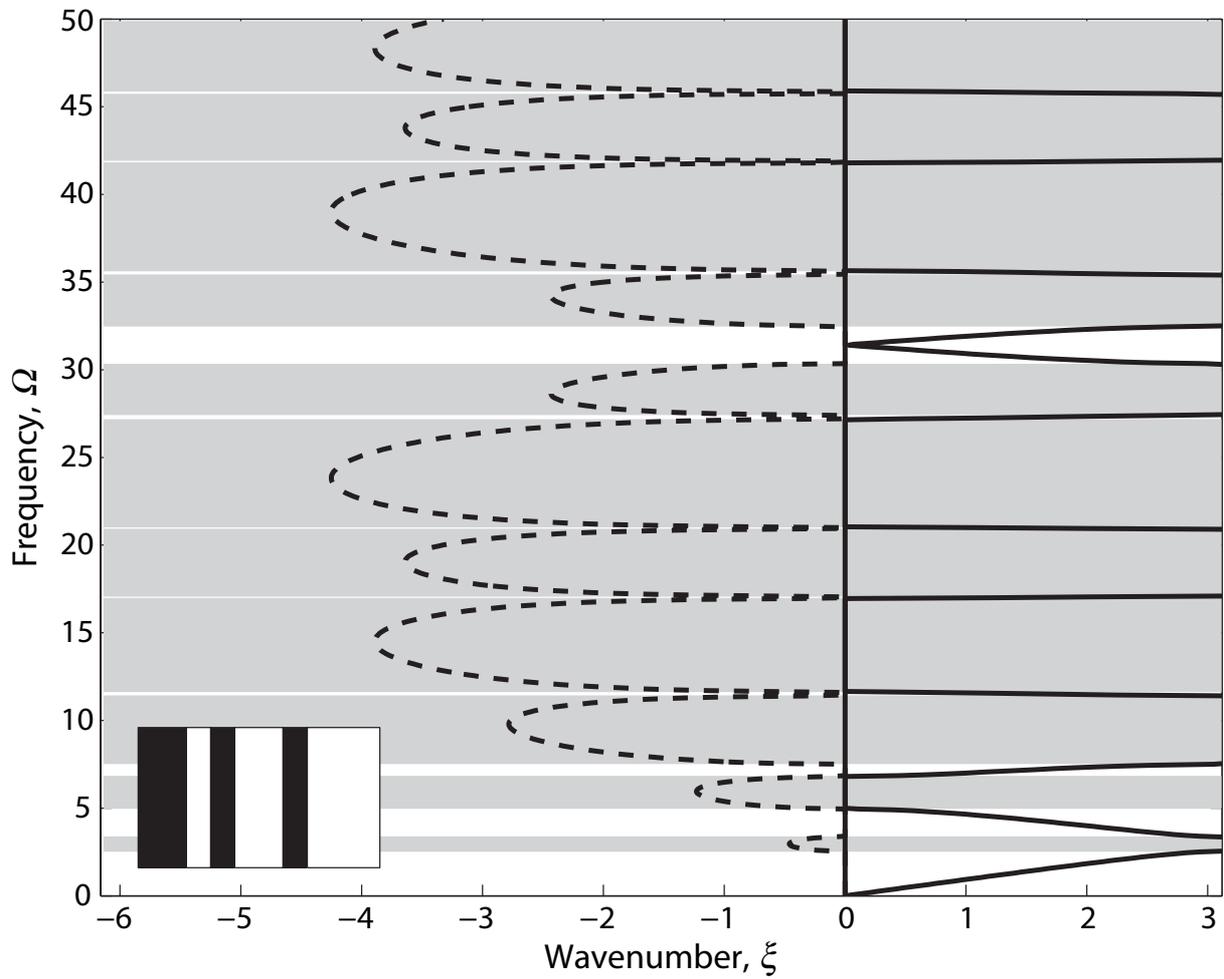

Figure 11　Frequency spectrum for optimal unit cell design for case involving general transient excitation at non-dimensional frequency range of $0 \leq \Omega^* \leq 50$. ( ——— ) Real part (pass band), ( – – ) Imaginary part (stop band). Stop bands are shaded. Design is shown in inset and has a binary code of '0010110111' (black: stiff material, white: compliant material).

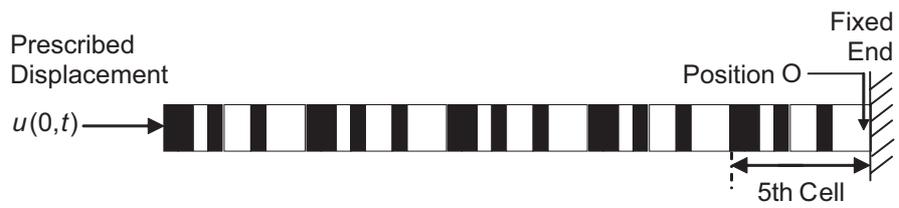

Figure 12  Banded structure designed for achieving shock isolation over frequency range $0 \leq \Omega^* \leq 50$. The structure is formed from the synthesized unit cell encoded '0010110111', shown in Fig. 2. Its length is $L = 5$. The excitation $u(0,t)$ is given by Eqs. (28)-(29). Position O marks a measurement point.

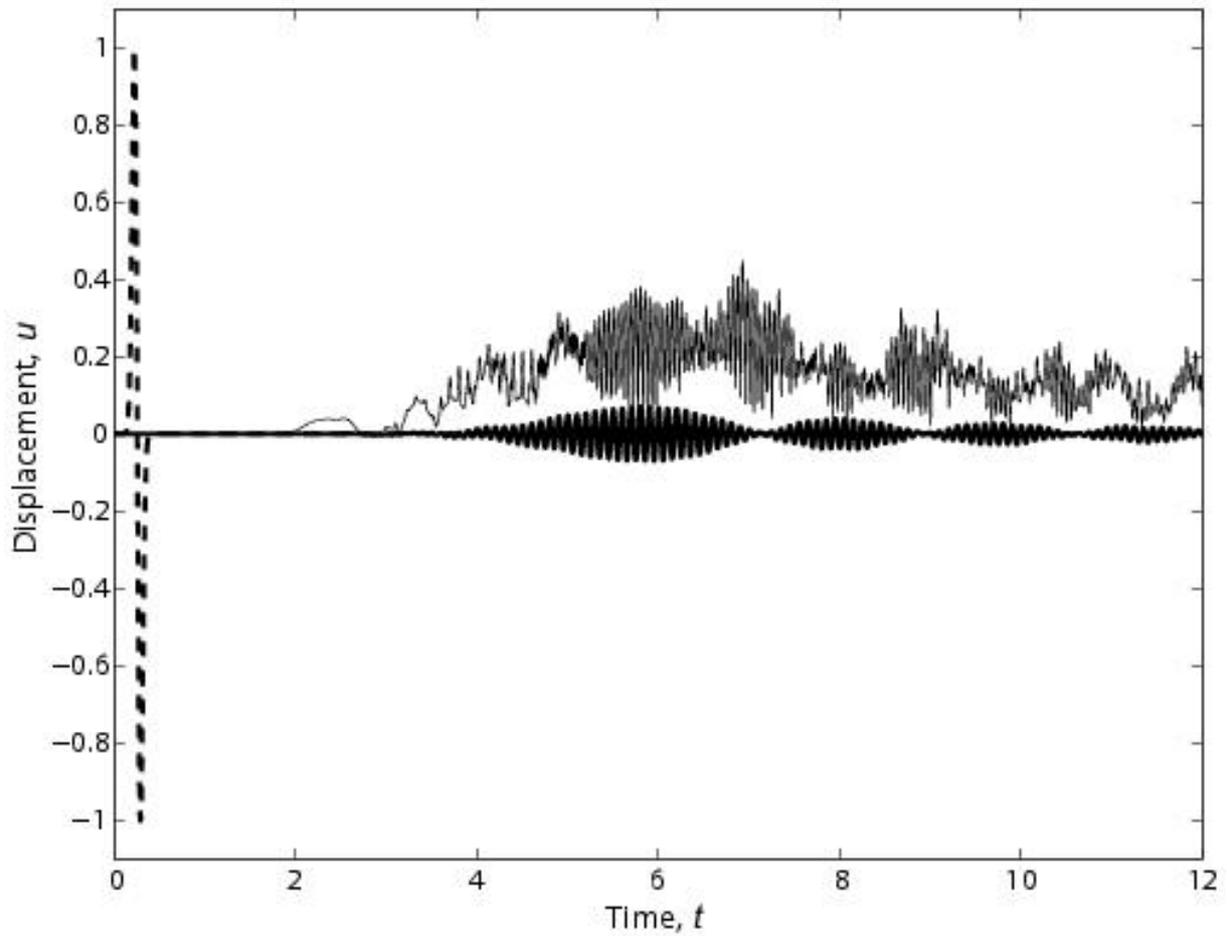

Figure 13 "Input" and "output" displacements for banded structure designed to isolate a shock pulse applied at the left end. The frequency content of the input pulse (given by Eqs. (28)-(29)) covers the range $0 \leq \Omega^* \leq 50$. ( ‒ ‒ ) Prescribed displacement at left end (input), (▬▬) Displacement close to right end at position O (output), (▬▬) Maximum displacement within 5th cell (which also represents the degree of transmission).

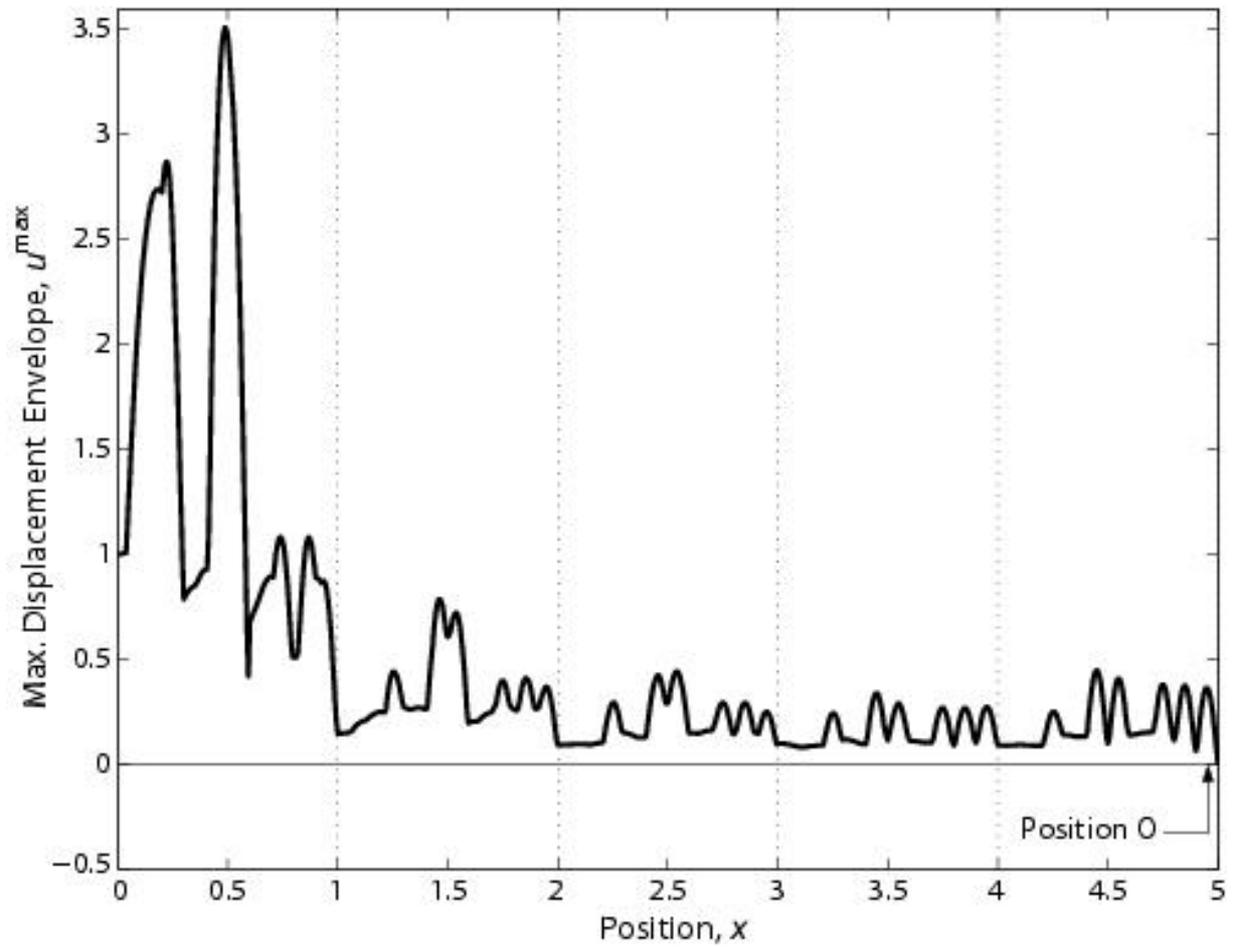

Figure 14  Envelope of maximum displacement throughout duration of numerical simulation (12 seconds) in banded structure designed for shock isolation. The shock load (given by Eqs. (28)-(29)), with frequency content spanning the range $0 \leq \Omega^* \leq 50$, is applied at the left end.

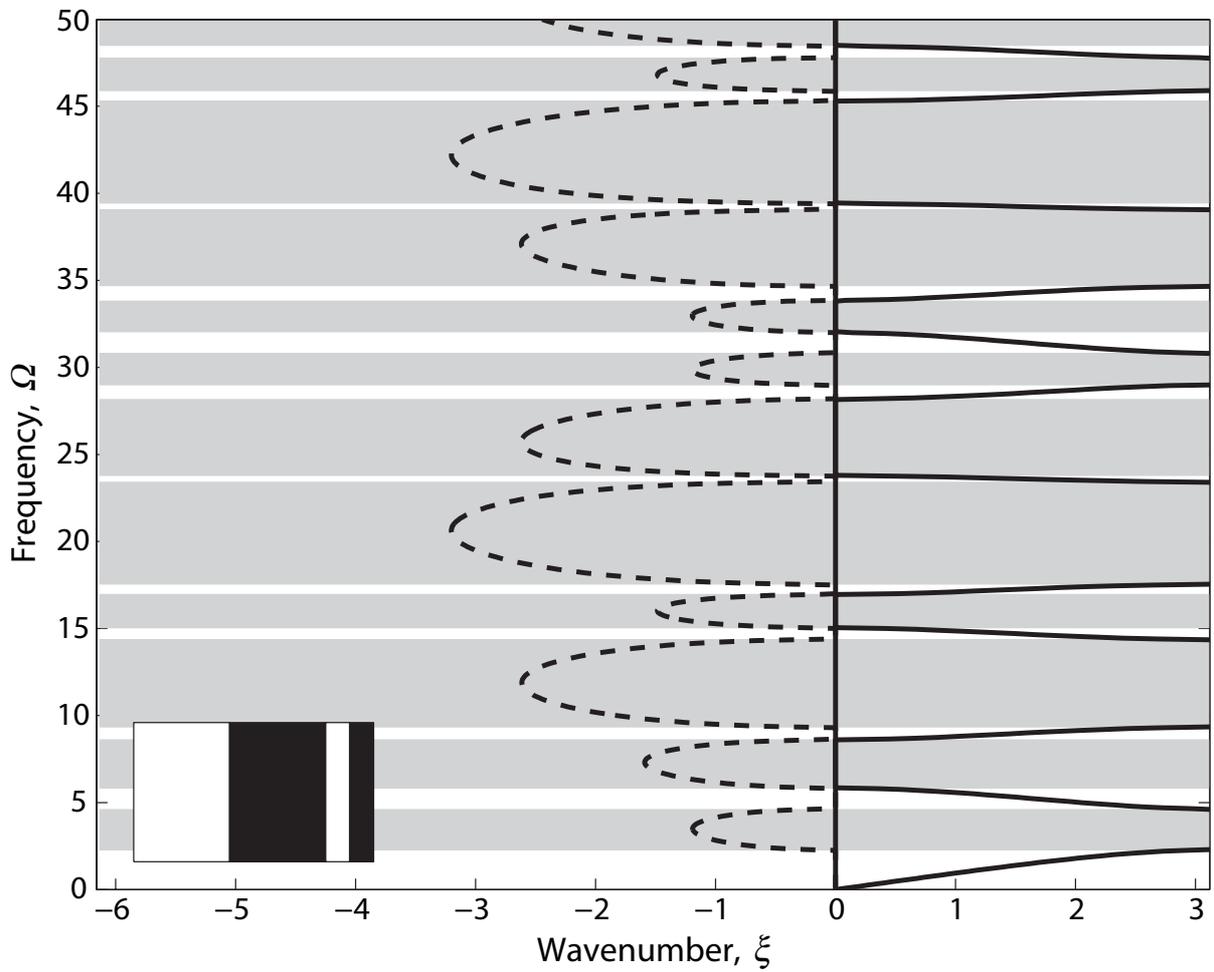

Figure 15  Frequency spectrum for unit cell designed to exhibit stop bands at frequency ranges that correspond to pass bands for the optimal unit cell of Fig. 11. (——) Real part (pass band), (− −) Imaginary part (stop band). Stop bands are shaded. Design is shown in inset and has a binary code of '1111000010' (black: stiff material, white: compliant material).

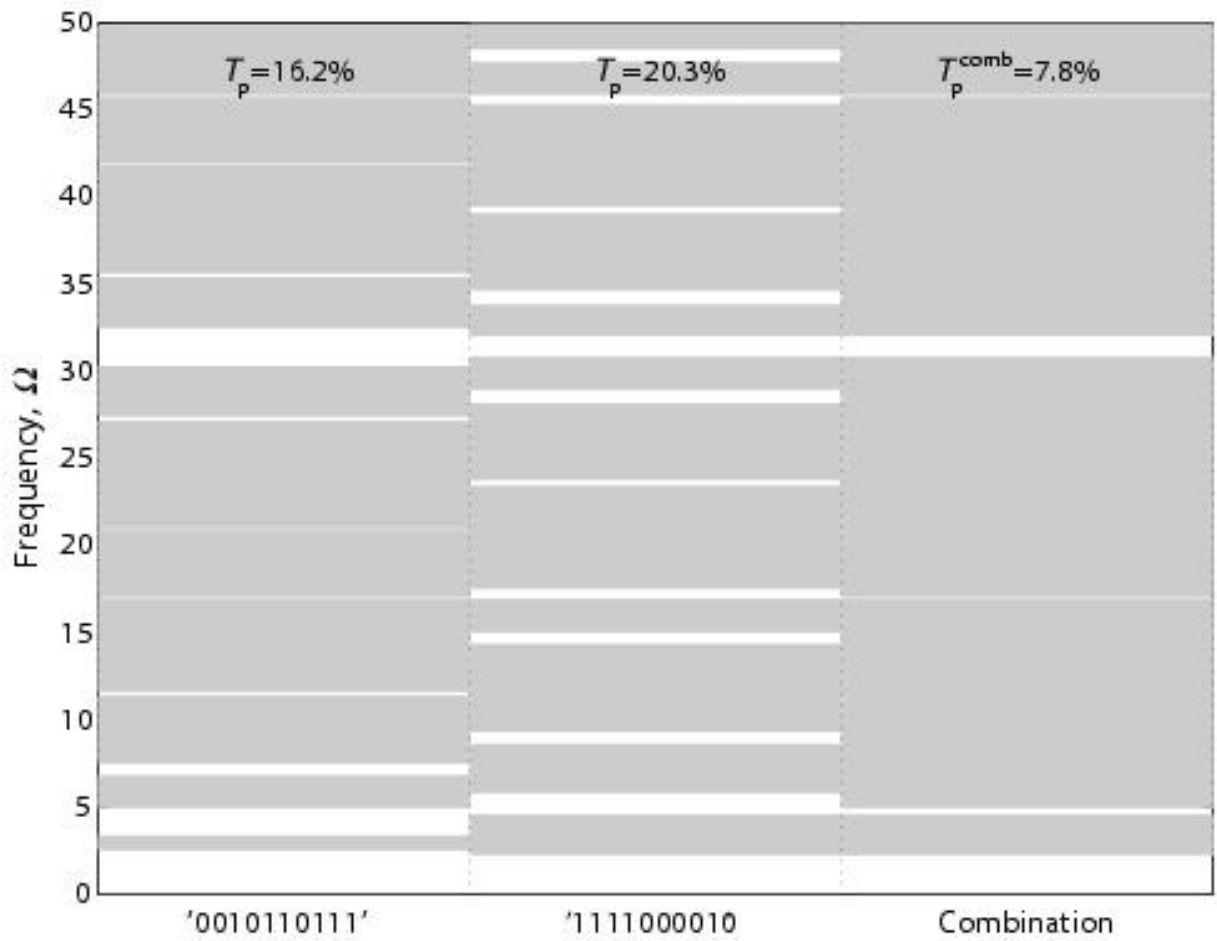

Figure 16  Schematic displaying band-gap frequency ranges (shaded) for optimal design encoded '0010110111' (Fig. 11), secondary design encoded '1111000010' (Fig. 15), and the two unit cells when viewed in "combination".

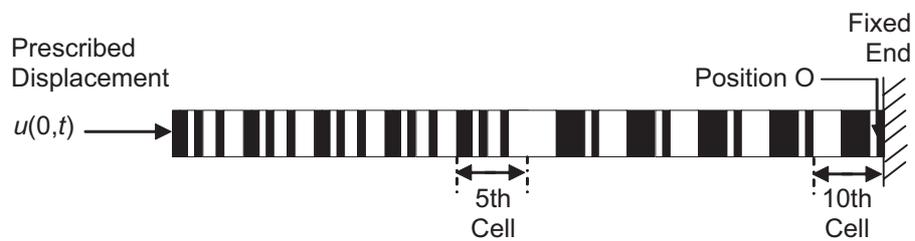

Figure 17  Banded structure designed for enhanced achievement of shock isolation over frequency range $0 \leq \Omega^* \leq 50$. The structure is formed from the synthesized unit cells '0010110111' (Fig. 11) and '1111000010' (Fig. 15). Five cells of the latter are stacked adjacent to five cells of the former. The structure's length is $L = 10$. The shock excitation $u(0,t)$ is given by Eqs. (28)-(29).

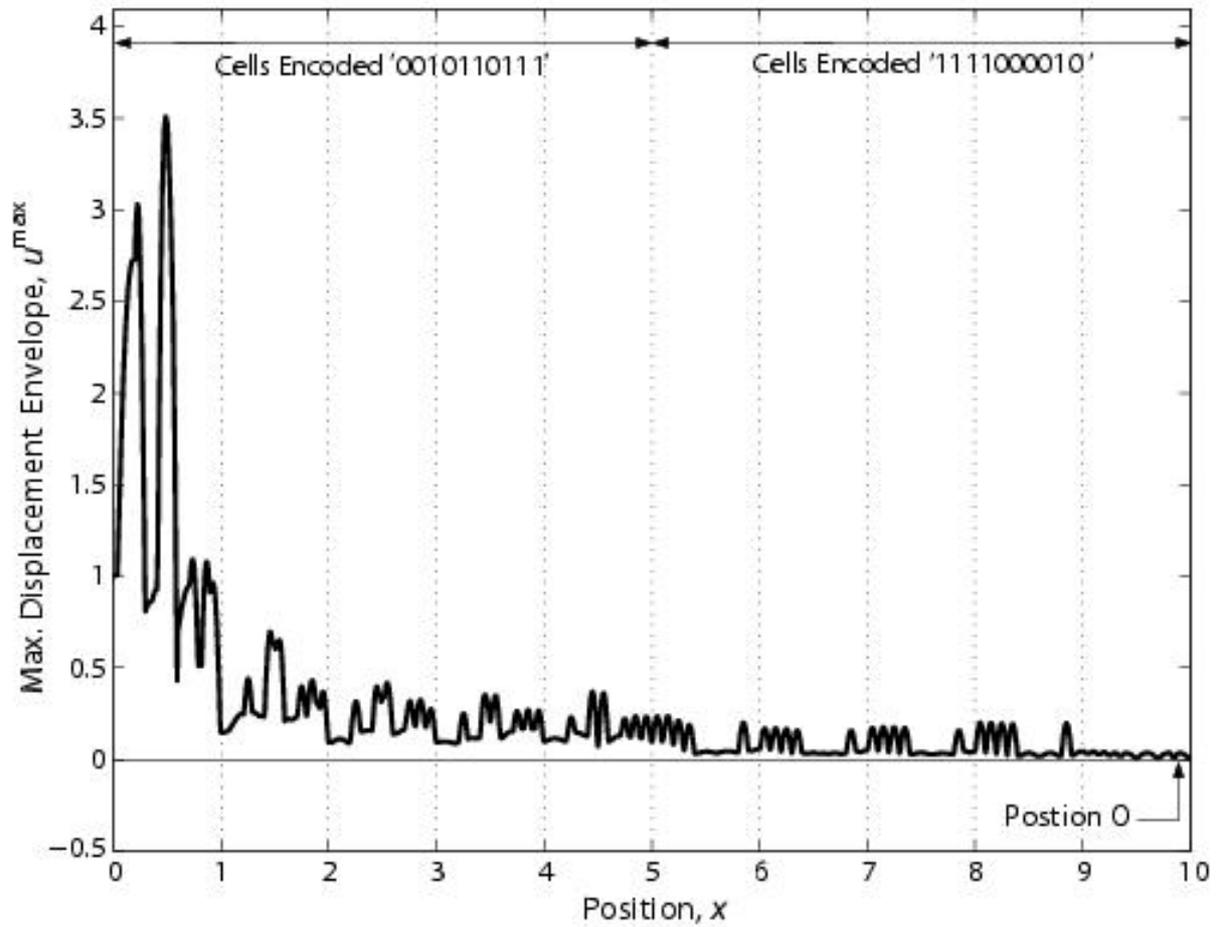

Figure 18　Envelope of maximum displacement throughout duration of numerical simulation (18 seconds) in multiple cell-type banded structure (top) and single cell-type banded structure (bottom) designed for shock isolation. The shock load (given by Eqs. (28)-(29)), with frequency content spanning the range $0 \leq \Omega^* \leq 50$, is applied at the left end.

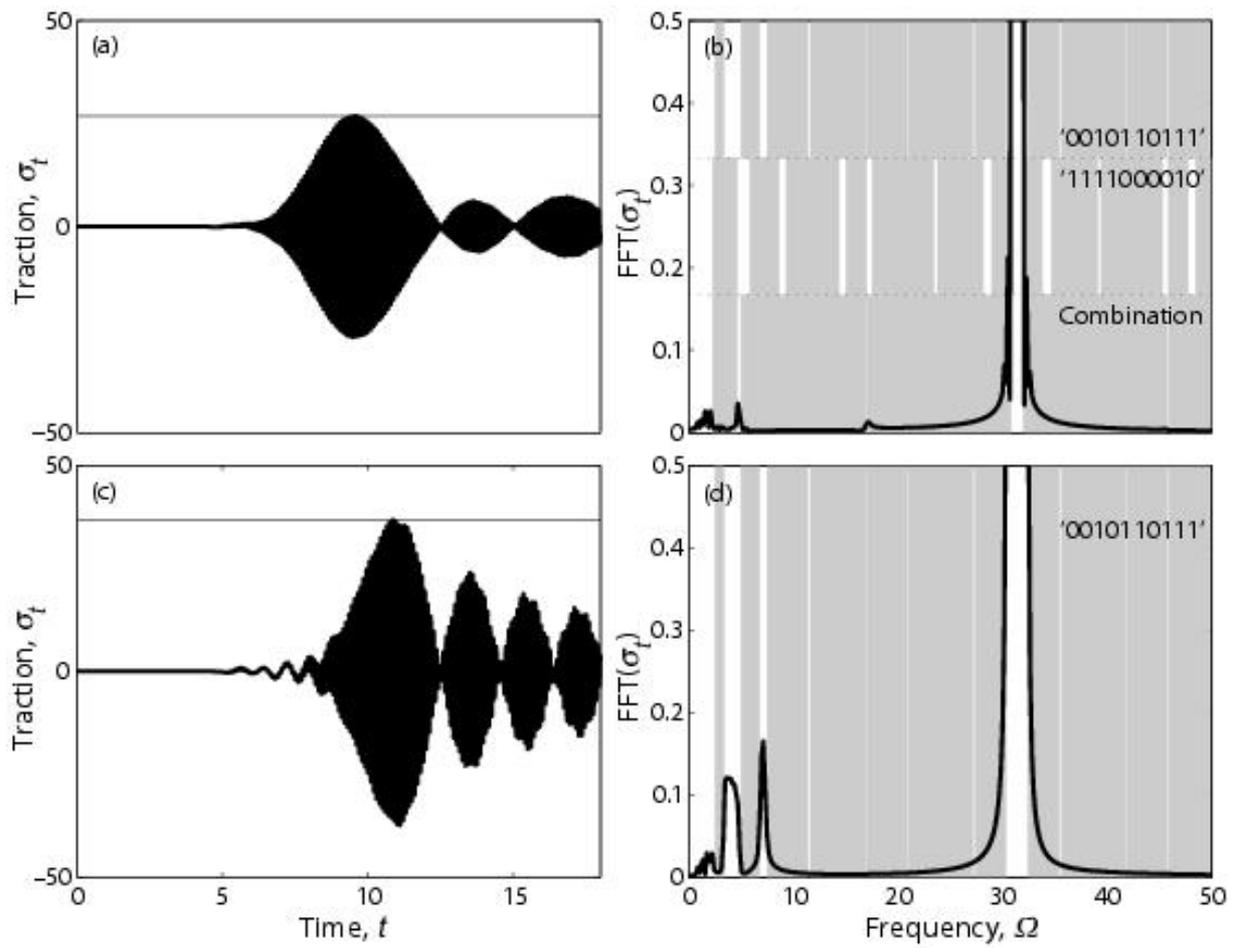

Figure 19 Transmitted traction at right (receiving) end for the multiple cell-type banded structure in the (a) time and (b) frequency domains. Traction at right (receiving) end for the single cell-type banded structure in the (c) time and (d) frequency domains. In the frequency domain plots, stop-band frequency ranges are shaded with appropriate labels.

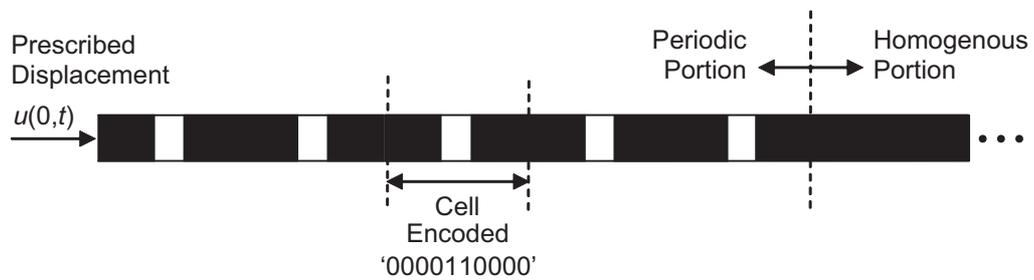

Figure 20   Structure consisting of five periodic cells of type '0000110000' and an extended homogenous portion. Loading in the form of prescribed harmonic displacement (starting at $t = 0$) at $\Omega^* = 5$ is applied at the left end.

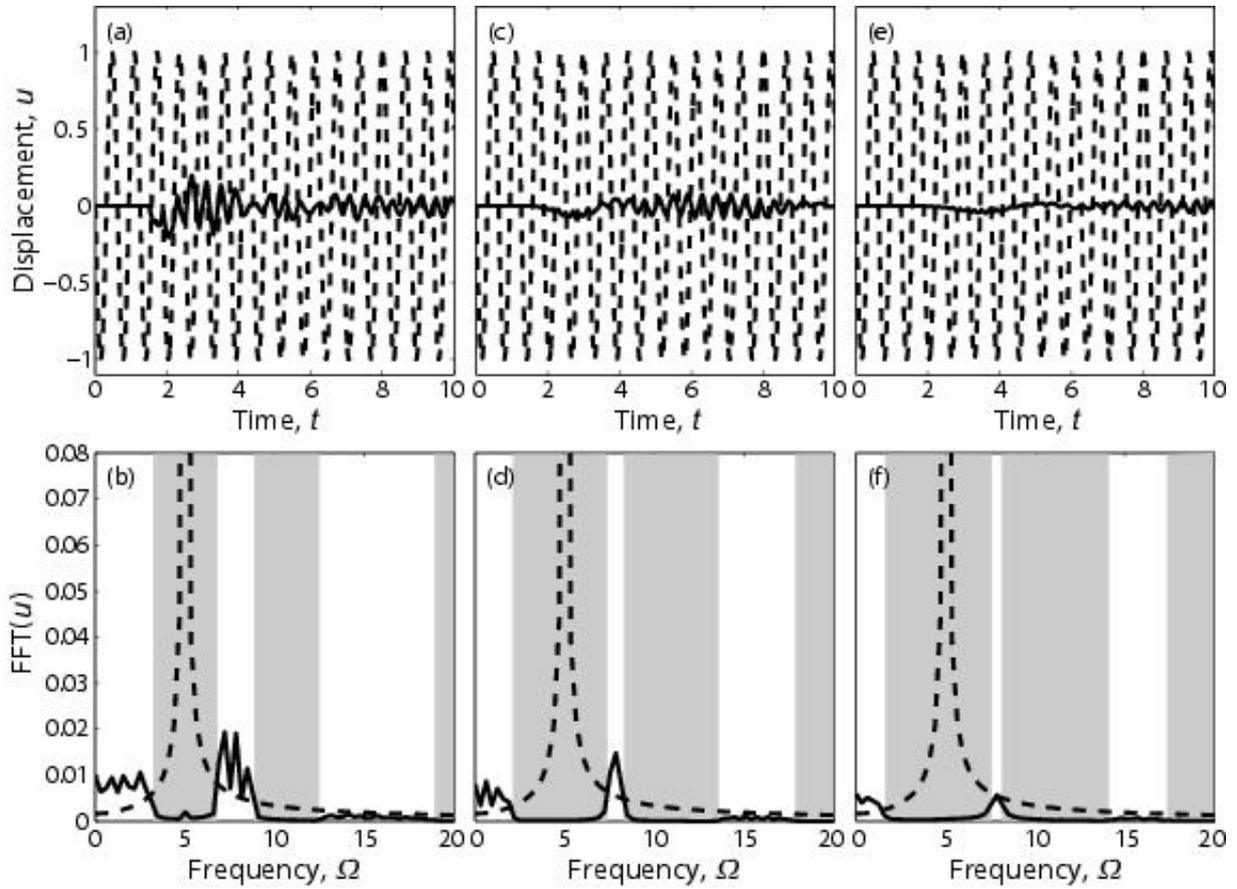

Figure 21 "Input" and "output" displacements in Fig. 20 structure for the following cases: $\rho_f/\rho_m=2$ in (a) time and (b) frequency domains, $\rho_f/\rho_m=5$ in (c) time and (d) frequency domains, and $\rho_f/\rho_m=9$ in (e) time and (f) frequency domains. ( **– –** ) Input at left end, ( —— ) Output at end of 5$^{th}$ cell. Stop-band frequency ranges for the employed unit cell are shaded in the frequency domain plots.

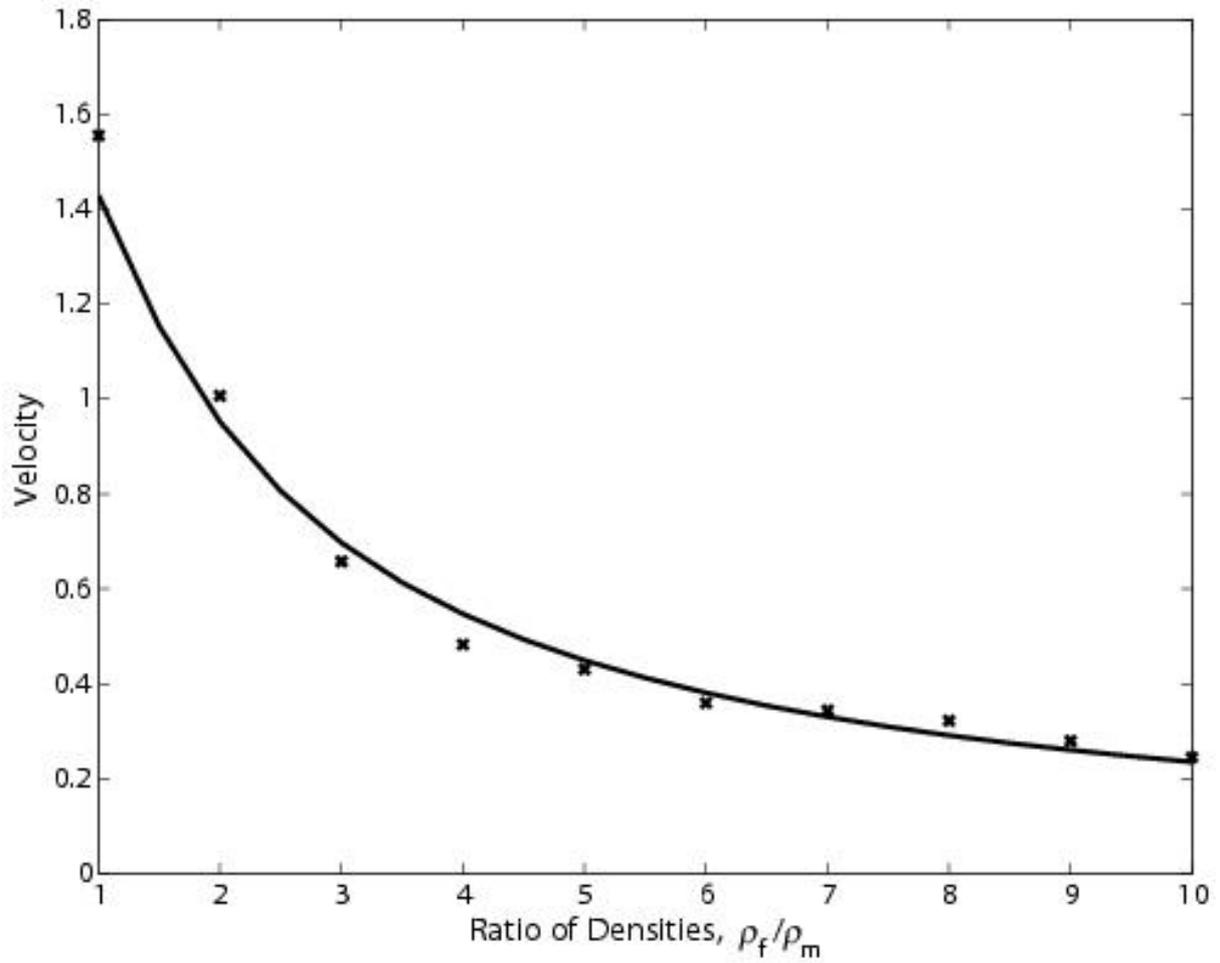

Figure 22  Velocity versus density ratio $\rho_f/\rho_m$ for both infinite media and a finite (five) number of cells (as shown in Fig. 20).  (———) Infinite: $\max(c_g)$ within 2nd pass band, (X) Finite: $\alpha$.

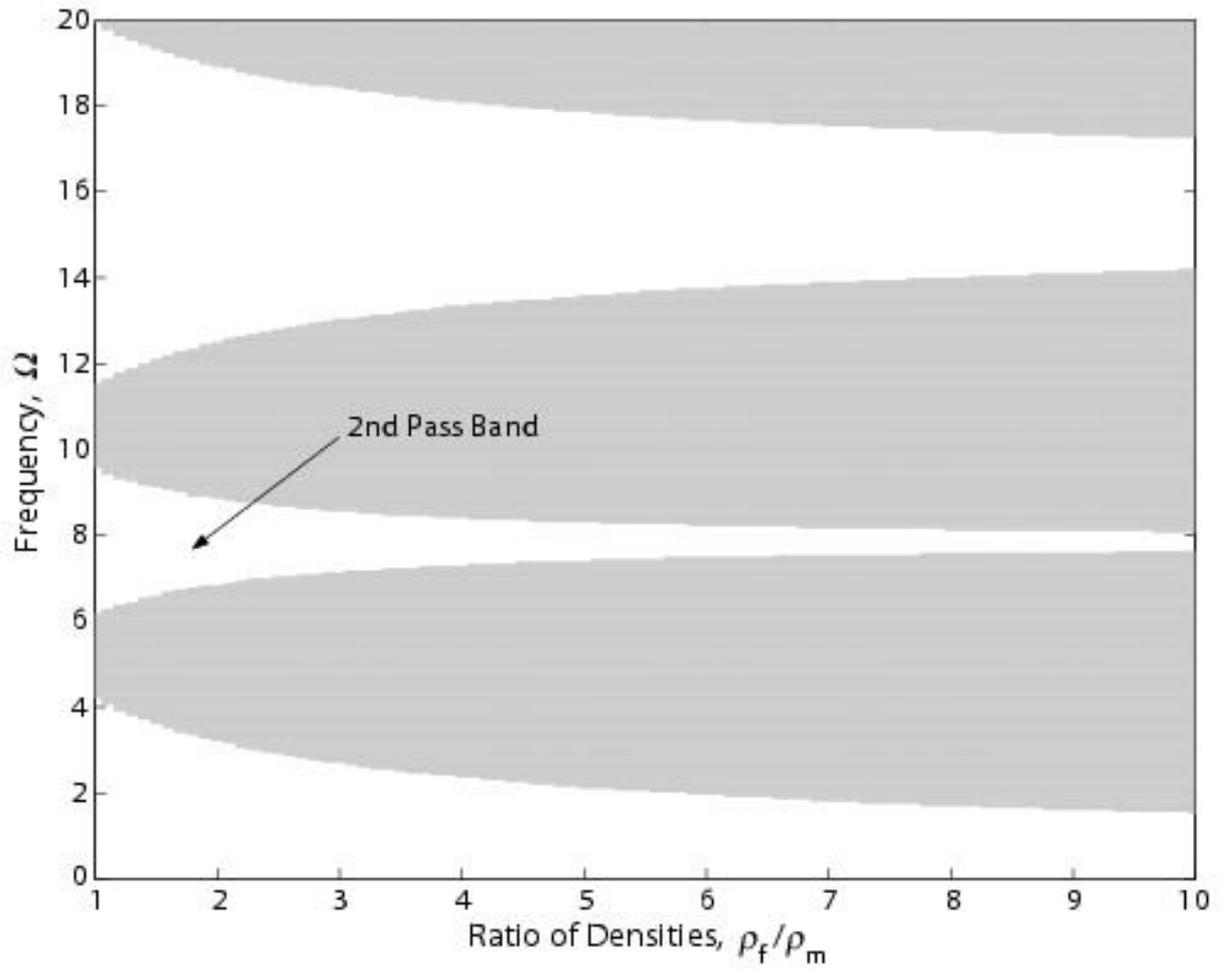

Figure 23    Frequency band-gap map for varying density ratio $\rho_f/\rho_m$ for unit cell encoded '0000110000'. Stop-band regions are shaded.

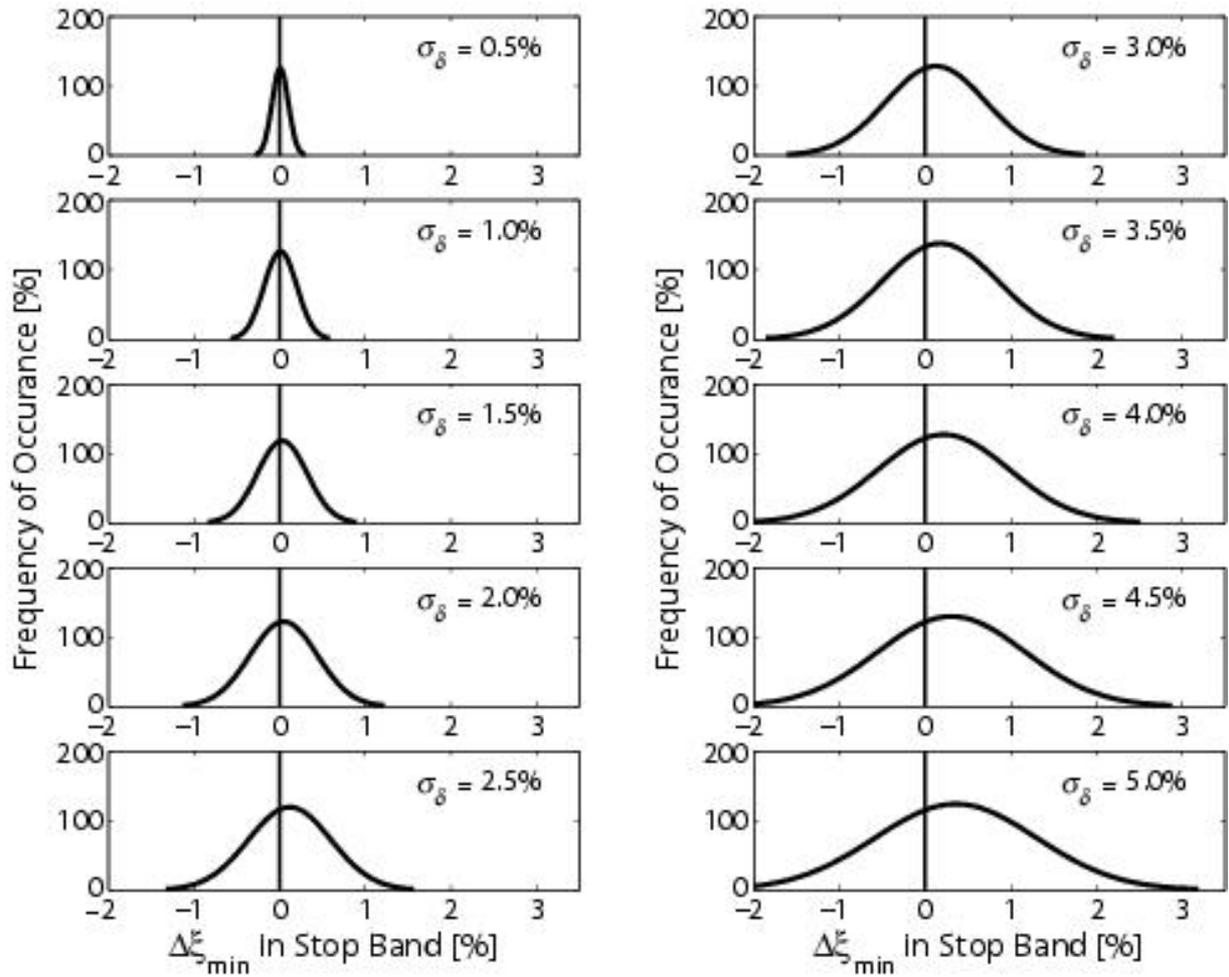

Figure 24   Sensitivity of frequency spectrum characteristics to statistical variations in layer thicknesses of unit cell '0101010101' (see Fig. 2). The change of the absolute value of the imaginary wavenumber at the center of the stop band of interest (i.e., at $\Omega^* = 20$) is denoted $\Delta\xi_{min}$.

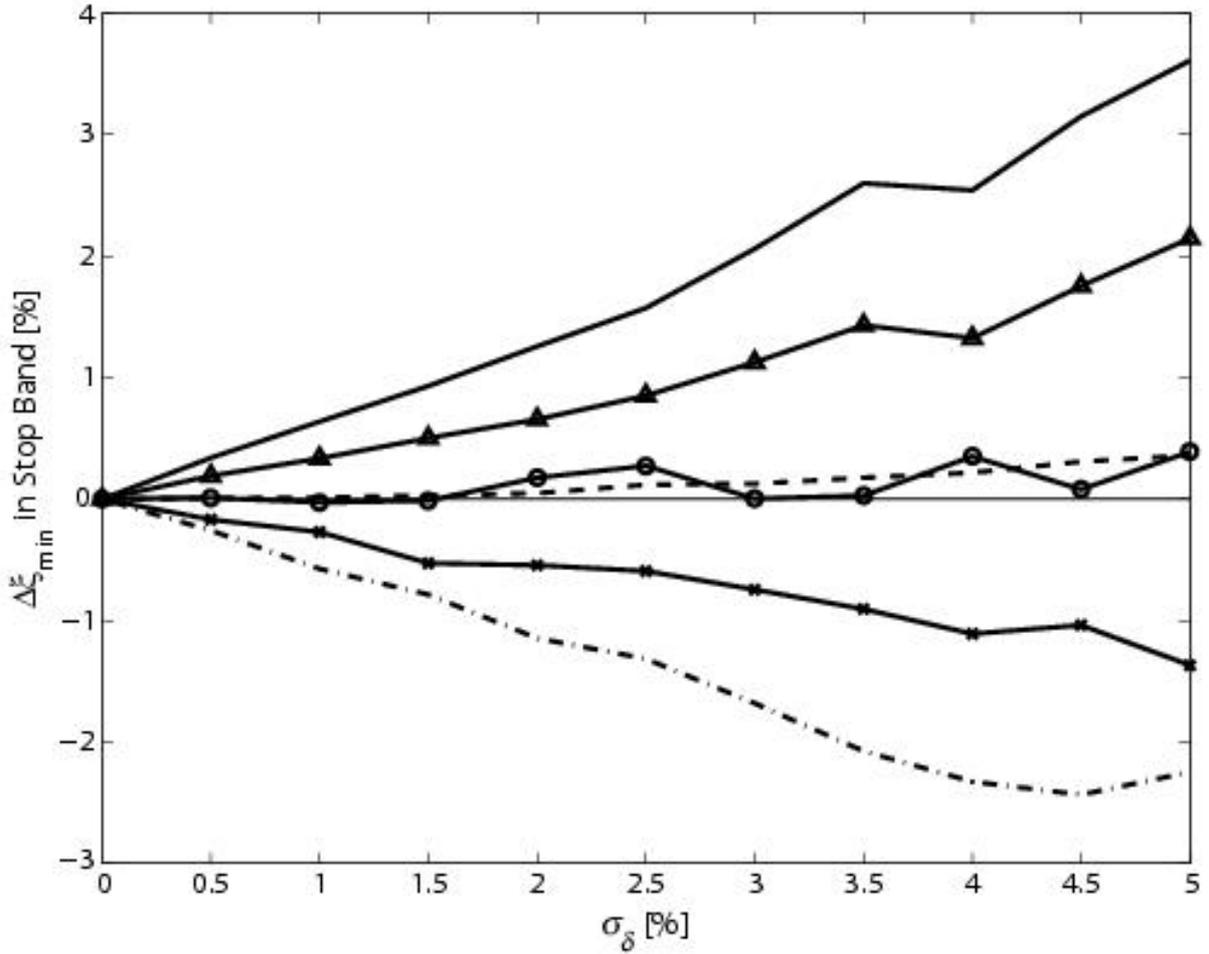

Figure 25　Sensitivity distribution quantities pertaining to the change of frequency spectrum characteristics (i.e., $\Delta\xi_{min}$ at $\Omega^* = 20$) as a result of statistical variations in layer thicknesses of unit cell '0101010101' (see Fig. 2). ( —— ) Maximum, ( − − ) Average, ( − · − ) Minimum, ( —△— ) 95th percentile, ( —○— ) 50th percentile, ( —✕— ) 5th percentile.

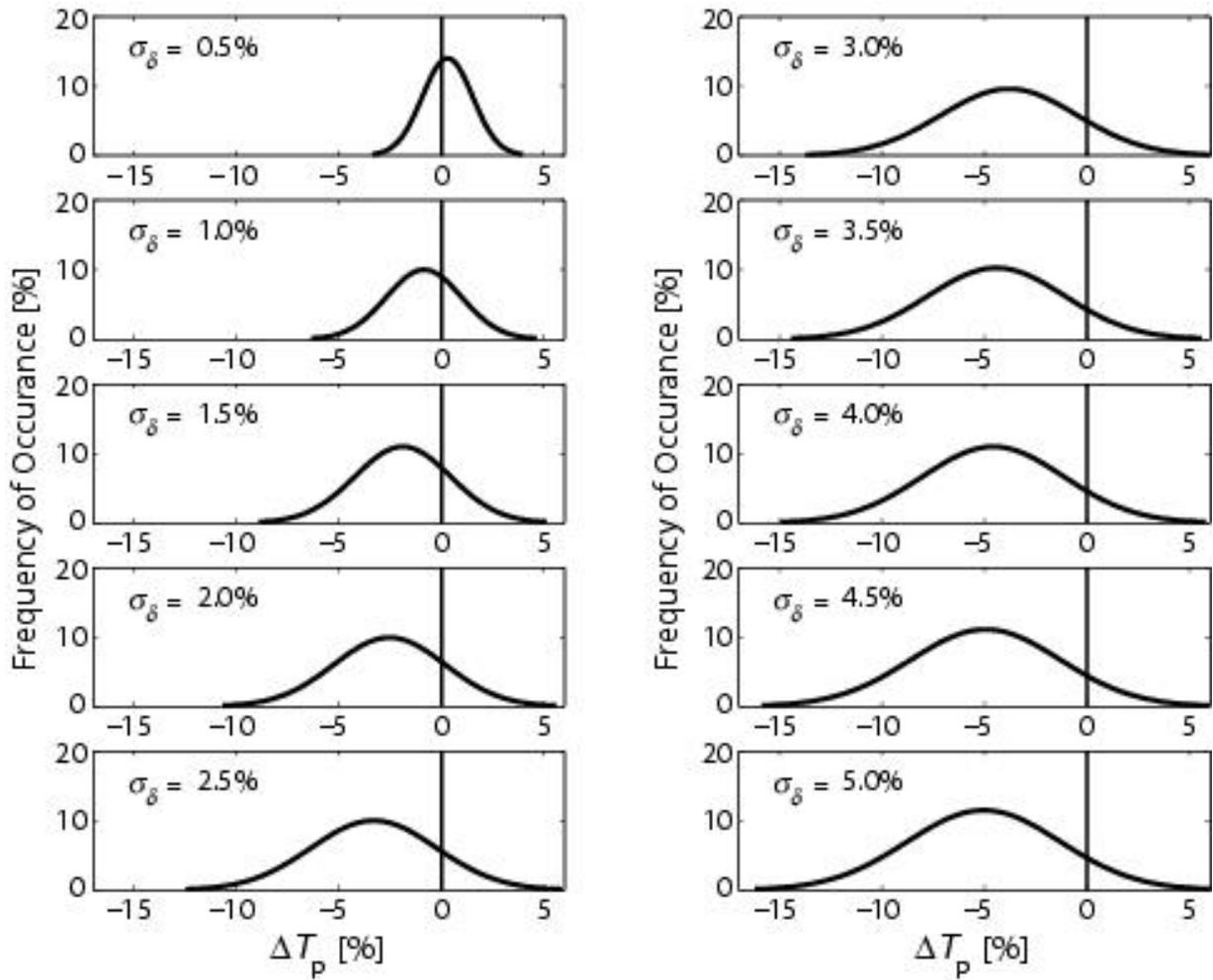

Figure 26  Sensitivity of frequency spectrum characteristics to statistical variations in layer thicknesses of unit cell '0010110111' (see Fig. 11). The change of the transmissibility within frequency range of interest (i.e., $0 \leq \Omega^* \leq 50$) is denoted $\Delta T_\mathrm{p}$.

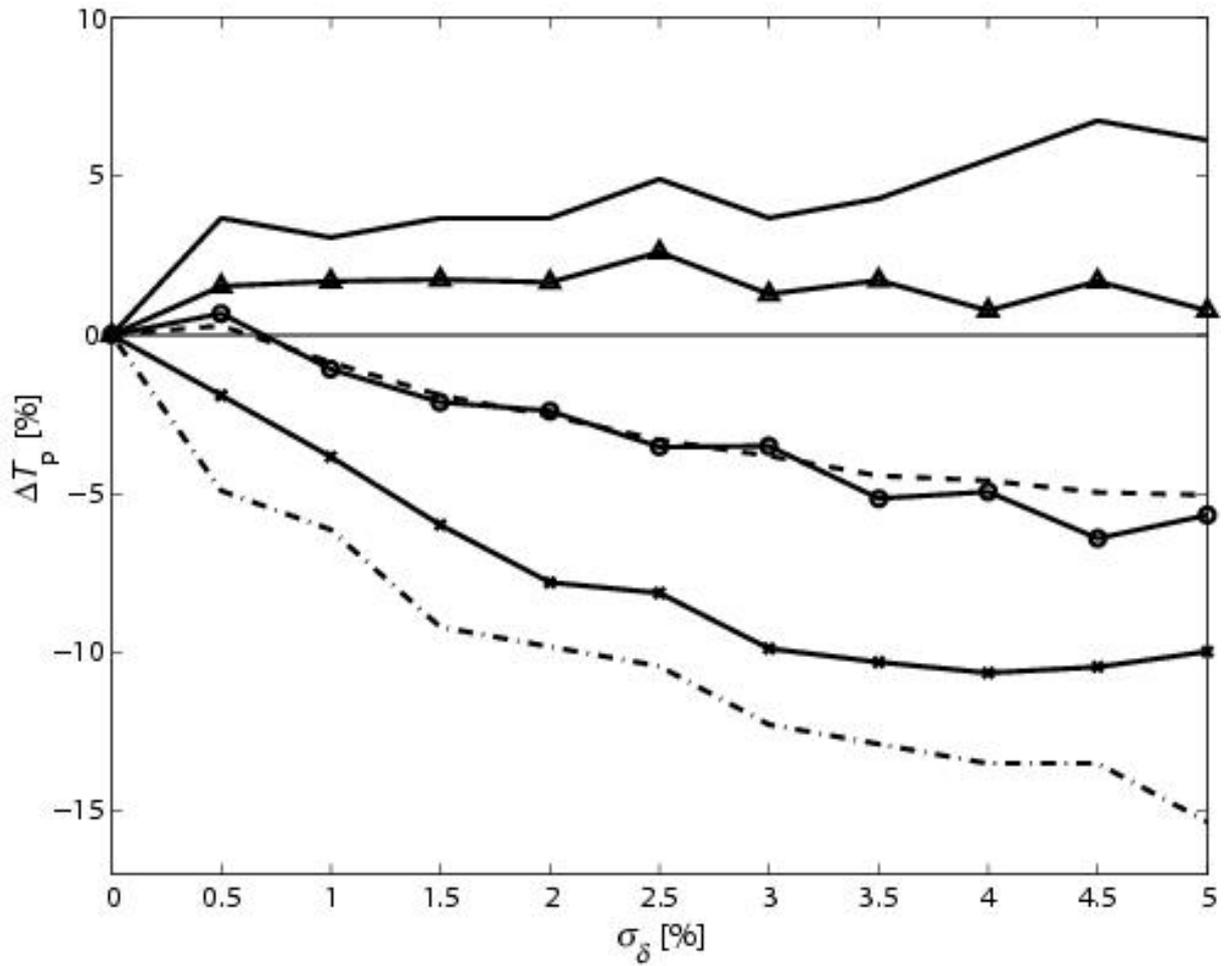

Figure 27  Sensitivity distribution quantities pertaining to the change of frequency spectrum characteristics (i.e., $\Delta T_p$ within $0 \leq \Omega^* \leq 50$) as a result of statistical variations in layer thicknesses of unit cell '0010110111' (see Fig. 11).  (———) Maximum, (— —) Average, (— · —) Minimum, (—▲—) 95th percentile, (—○—) 50th percentile, (—✕—) 5th percentile.